\documentclass[%
amsmath,amssymb,
]{revtex4-2}
\usepackage{graphicx}
\usepackage{dcolumn}
\usepackage{bm}
\usepackage{hyperref}
\usepackage{lineno}
\usepackage{url}
\usepackage{xcolor}
\usepackage{titlesec}
\usepackage[singlelinecheck=false,justification=raggedright]{caption}
\usepackage[labelformat=simple]{subcaption}

\usepackage{multirow}
\usepackage{ulem}
\usepackage{cancel}
\usepackage{xcolor}

\titlespacing*{\section}
{0pt}{4ex}{2ex}
\titlespacing*{\subsection}
{0pt}{4ex}{2ex}
\titlespacing*{\subsubsection}
{0pt}{4ex}{2ex}

\begin{document}
\title{Diffusion-based sensor placement optimization and reconstruction of turbulent flows in urban environments}

\author{Abhijeet Vishwasrao$^{\text{1,2,}*}$, Sai Bharath Chandra Gutha$^{\text{3}}$,  Andres Cremades$^{\text{2,4}}$, Klas Wijk$^{\text{3,5}}$, Aakash Patil$^{\text{6,7}}$, H.D. Lim$^{\text{8}}$,
Christina Vanderwel$^{\text{9}}$,  Catherine Gorle$^{\text{10}}$, Beverley J McKeon$^{\text{6}}$, Hossein Azizpour$^{\text{3,5}}$,  Ricardo Vinuesa$^{\text{1,2,}*}$}

\affiliation{1: Department of Aerospace Engineering, University of Michigan, Ann Arbor, USA \\
2: FLOW, Engineering Mechanics, KTH Royal Institute of Technology, Stockholm, Sweden \\
3: Division of Robotics, Perception and Learning, School of EECS, KTH Royal Institute of Technology, Stockholm, Sweden \\
4: Instituto Universitario de Matemática Pura y Aplicada, Universitat Politècnica de València, Spain\\
5: Swedish e-Science Research Centre (SeRC) \\
6: Department of Mechanical Engineering and Center for Turbulence Research, Stanford University, USA \\
7: Nugen Intelligence, Mumbai, India \\
8: Temasek Laboratories, National University of Singapore, Singapore \\
9: Department of Aeronautics and Astronautics, University of Southampton, Southampton, United Kingdom \\
10: Department of Civil Engineering, Stanford University, USA \\
$^{*}$Corresponding authors: abvish@umich.edu; rvinuesa@umich.edu \\
}%


\begin{abstract}

Rapid urbanization demands efficient monitoring of turbulent wind and pollutant dispersion, yet existing reconstruction and sensor placement strategies fail under realistic sparsity constraints. Here, we introduce Diff--SPORT, a diffusion-based framework that combines a generative diffusion prior with maximum a posteriori inference and Shapley-value attribution for high-fidelity flow reconstruction and optimal sensor placement. By training a diffusion prior model once over a domain, Diff--SPORT enables non-linear optimal sensor placement and near-real-time flow reconstruction from sparse measurements orders of magnitude faster than RANS or LES simulations, consistently outperforming state-of-the-art methods. The framework also extends, without algorithmic modification, to experimental passive scalar concentration dataset, a direct proxy for pollutant dispersion, measured in a 1:2400 scale water-flume model of the Beijing Haidian neighbourhood under realistic urban flow conditions. Shapley-guided sensor placement achieves up to 57\% lower reconstruction error than randomly placed sensors at extreme sparsity, identifying compact and physically interpretable configurations. These results establish Diff--SPORT as a modular foundation offering a zero-shot alternative to retraining-intensive downstream strategies, supporting scalable urban flow monitoring for air quality management and resilient city design.

\end{abstract}

\maketitle

\section*{Introduction}
Monitoring turbulent flows in urban areas is a cornerstone of sustainable city design, crucial for ensuring air quality~\citep{Zhang2013}, mitigating urban heat island effect~\citep{Manoli2019, Sen2021} and enabling responsive infrastructure~\citep{Balaian2024, Sen2021}. Turbulent wind patterns govern key processes in cities, such as pollutant dispersion~\citep{Zhang2013} (linked to 7 million annual premature deaths attributed to air pollution~\citep{who2014}), pedestrian comfort~\citep{Sen2021}, microclimate regulation~\citep{VinuesaReview}, to name a few. With increasing urbanization and climate-induced stressors, the need for a real-time, accurate reconstruction of these flows has become a practical necessity. Efficient modeling of these dynamics supports resilient city planning~\citep{Zhang2013, Balaian2024, Sen2021} and enhances the reliability of emerging technologies operating in urban environments. In particular, accurate flow reconstruction through optimally placed sensor measurements is essential for anticipating pollution hot-spots, mitigating wind-induced stress and adapting to extreme weather events.

Since fluid flows are governed by the Navier--Stokes equations, a set of nonlinear partial differential equations (PDEs), numerical methods such as Reynolds-averaged Navier--Stokes (RANS), large-eddy simulation (LES) and direct numerical simulation (DNS) are commonly employed to study complex wind patterns, depending on the required spatiotemporal flow resolution~\citep{Li2025}. RANS yields low-fidelity, computationally efficient flow fields, while LES and DNS provide high-resolution data capturing multiscale turbulent dynamics. However, their high computational cost renders real-time wind field monitoring impractical with current resources, despite their superior accuracy. To address this limitation, sparse reconstruction techniques have been developed. These methods rely on limited sensor measurements deployed within urban canopies to reconstruct flow fields utilizing data assimilation techniques. To this end, there have been efforts to integrate measurements directly into the numerical schemes~\citep{gu_application_2017} and methods based on regressive reconstruction using stochastic estimation~\citep{perret_combining_2016, dubois_machine_2022} among others, have been developed. Another widely utilized data-driven strategy involves using linear combinations of reference modes, which can be derived from several modal analysis techniques~\citep{taira2017modal}. For example, Gappy-POD, where the best linear combination of proper orthogonal decomposition (POD) modes is determined by solving a least-squares problem, has been successfully applied to reconstruct three-dimensional (3D) flows past a cavity with a similar number of modes and sensors~\citep{pastur_pod-based_2008}. However, these approaches are often tailored to canonical flows and are limited when dealing with turbulent flows, as the number of spatial modes required for accurate reconstruction becomes prohibitively large.

Recent advancements in convolutional-neural-network (CNN)-based architectures, have shown significant potential in improving image inpainting and super-resolution for natural images. Building on these developments, supervised deep learning methods, such as generative adversarial networks (GANs)~\citep{goodfellowGenerativeAdversarialNetworks2014}, autoencoders (AEs)~\citep{Kramer1991NonlinearPC}, variational autoencoders (VAEs)~\citep{kingmaAutoEncodingVariationalBayes2014} and transformers have been extensively applied for sparse reconstruction, super-resolution or flow prediction tasks for fluid flows~\citep{li_generative_2023, dubois_machine_2022, Solera-Rico2024, EIVAZI2022117038}. Although these methods outperform traditional interpolations (e.g., nearest neighbors, bicubic) at capturing large-scale flow structures, their deterministic nature hinders generalization to finer scales and restricts applicability to canonical flow scenarios~\citep{li_generative_2023, chuang_experience_2022}. 

To this end, diffusion models have emerged as promising tools for learning complex, high-dimensional data distributions and have shown strong performance towards generating synthetic turbulent data for both the Lagrangian~\citep{li_synthetic_2024} and Eulerian~\citep{shu_physics-informed_2023, zhuang_spatially-aware_2024, du_confild_2024, Li_S3GM_2024, Vishwasrao2024} frames of reference. A straightforward approach for sparse reconstruction or super-resolution with diffusion models is to directly map the sensor measurements to the flow field~\citep{zhuang_spatially-aware_2024, shu_physics-informed_2023}, however, this approach does not fully utilize the stochastic potential of the diffusion models. 
A more nuanced line of work has recently emerged, using diffusion models as probabilistic priors for flow inference. \citet{gao_bayesian_2024} train an unconditional spatiotemporal diffusion model and introduce downstream conditioning via global physics-based constraints to guide the reverse diffusion process; while effective for dense, globally informative observations, this approach does not extend naturally to sparse-sensor reconstruction, where likelihood information is highly localized. More recently, \citet{du_confild_2024} and related studies~\citep{Li_S3GM_2024, Vishwasrao2024} advanced this direction by demonstrating zero-shot sparse-sensor reconstruction using score-based conditional sampling methods such as DDRM~\citep{kawarDenoisingDiffusionRestoration2022}, $\Pi$GDM~\citep{songPseudoinverseGuidedDiffusionModels2022}, and DPS~\citep{chung2023diffusion}. However, as sparsity increases, the gradient signal from the likelihood term becomes increasingly weak and localized, limiting these methods' ability to steer the generative dynamics and leading to performance degradation under tight sensor budgets. Moreover, these approaches rely on spatiotemporal priors trained on sequential flow data, implicitly requiring temporally synchronized and regularly sampled observations at inference, a condition rarely satisfied in operational urban environments, where sensor measurements arrive asynchronously. These limitations highlight the need for more robust, adaptable approaches that can handle sparse data, turbulent stochasticity and real-world constraints on sensor placement.

It is imperative to note that the sparse reconstruction problem is inherently coupled with the optimal sensor placement. Optimal sensor placement, or more broadly, feature selection, involves identifying a subset of the most relevant features from a larger dataset to maximize the accuracy and efficiency of sparse reconstruction. For urban flow applications, the literature broadly categorizes sensor placement strategies into two main approaches: mode-decomposition methods and deep-learning-based methods. Similar to sparse reconstruction techniques, methods based on POD like QR-pivoting~\citep{Manohar2017DataDrivenSS} and multi-resolution dynamic mode decomposition (mrDMD)~\citep{kelp_datadriven_2023} have been widely utilized. These techniques use POD modes derived from velocity or vorticity fields as basis functions to obtain a latent representation of flow fields. While interpretable and effective in many applications, POD-based methods are limited by their linear nature and the computational challenges associated with performing the singular value decomposition (SVD) on large-scale 3D data, which is typical in urban environments, paving the way for more advanced deep-learning based techniques.

The success of deep-learning has catalyzed the development of neural network-based embedded feature selection methods, which show significant promise for problems with many degrees of freedom (DOFs)~\citep{balinConcreteAutoencodersDifferentiable2019, huijbenDeepProbabilisticSubsampling2019, yamadaFeatureSelectionUsing2020}. Concrete autoencoders (CAEs)~\citep{balinConcreteAutoencodersDifferentiable2019} and indirectly parameterized concrete autoencoders (IP-CAEs)~\citep{nilssonIndirectlyParameterizedConcrete2024}, for instance, provide an end-to-end differentiable framework that selects the most informative features from input data. Another such approach, based on attention-backed AEs, is Senseiver~\citep{santos_development_2023}. While these methods outperform linear approaches in identifying optimal sensor locations, they lack the flexibility and modularity needed for practical deployment, requiring retraining for minor workflow changes. In urban settings, sensors must be placed in accessible areas, near walls or the ground, without disrupting activities or altering flow fields. Constraints like installation cost, power efficiency, and limited on-device computing often demand fewer but larger sensors, revealing a significant gap between theoretical sensor placement strategies and real-world feasibility.

To address these challenges, we propose Diff--SPORT: a diffusion-based framework for Sensor Placement Optimization and Reconstruction of Turbulence. Diff--SPORT combines a generative diffusion model trained on DNS data with (i) a gradient-based MAP inference algorithm~\citep{Gutha2024InversePW} for sparse reconstruction and (ii) a Shapley-value-based~\citep{shapley1953value} feature attribution framework for optimal sensor placement. The core idea is to use a trained diffusion model as a probabilistic surrogate or a foundation model~\citep{Bommasani2021} for the flow-field distribution, enabling both unconditional generation and posterior-based reconstruction. Our maximum a posteriori gradient ascent (MAPGA) method directly optimizes the posterior $p_\theta(\mathbf{\Psi}^{\prime}_0 | \mathcal{S})$ given sparse sensor data $\mathcal{S}$, leveraging the differentiability of the diffusion prior and avoiding the approximations required by score-based approaches such as $\Pi$GDM~\citep{songPseudoinverseGuidedDiffusionModels2022}. For sensor placement, we adapt the game theory-based shapley additive explanations (SHAP) framework~\citep{Lundberg} to probabilistic settings by introducing a custom weighting kernel that emphasizes coalitions within an optimal range. This enables scalable and interpretable ranking of candidate sensor subregions based on their contribution to reconstruction accuracy.

 Building on recent advances in zero-shot diffusion-prior models for turbulent flows, Diff--SPORT adopts a spatial IID prior trained on independent snapshots, enabling zero-shot inference from any instantaneous observation without temporal synchronization requirements, a deliberate design choice for asynchronous urban sensor networks (see Discussion), and contributes to two methodological innovations. First, a scalable MAP-based sparse-reconstruction framework that directly optimizes the posterior without relying on score approximations, achieving superior accuracy over state-of-the-art score-based methods ($\Pi$GDM, DPS) particularly under extreme sparsity and additional constraints. Second, the integration of Shapley value attribution with diffusion-based posterior inference to identify deployable, high-utility sensor locations, the first framework to combine game-theoretic feature attribution with probabilistic generative priors for flow reconstruction.  Validated on both a high-fidelity DNS benchmark of a single-building geometry and an experimental passive scalar dataset from a 14-building neighbourhood at $\mathrm{Re} \approx 14{,}000$, Diff--SPORT combines statistical fidelity with practical deployability, advancing interpretable, generative approaches for sustainable and smart urban infrastructure.

\section*{Results}

\subsection*{Overview of Diff--SPORT framework}
\begin{figure}
    \begin{center}
	    	\includegraphics[width=1\textwidth]{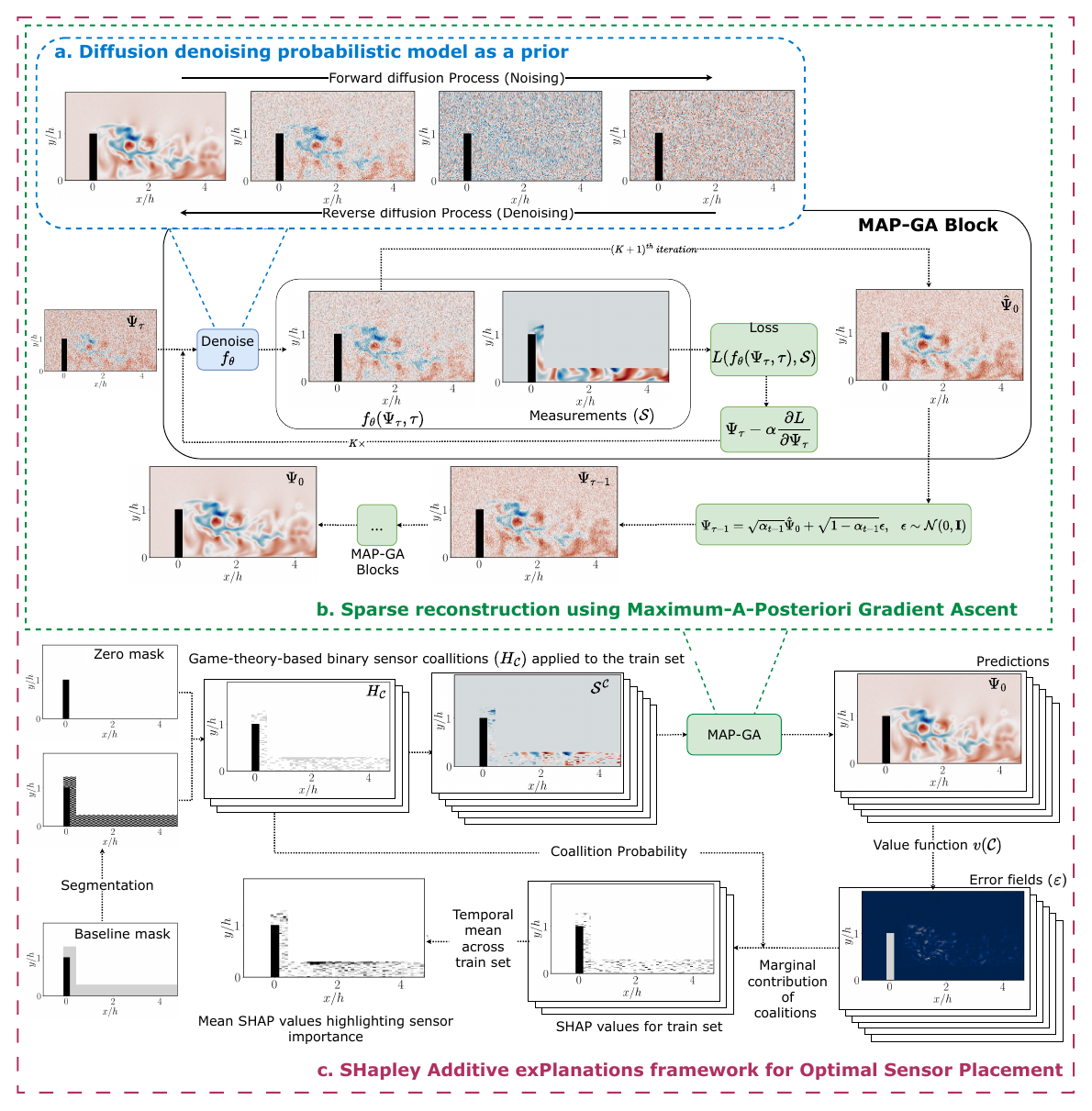}
        	\caption{\label{fig:overview}
            \textbf{Overview of the Diff--SPORT framework for sparse flow reconstruction and sensor placement.} 
            (a) A denoising diffusion probabilistic model (DDPM) is trained as a generative prior for urban turbulent flow simulation. The forward process adds Gaussian noise to an initial velocity-fluctuation field \( \mathbf{\Psi}^{\prime}_0 \) to produce increasingly degraded representations \( \mathbf{\Psi}^{\prime}_\tau \), while the reverse process iteratively denoises back to a plausible sample from the learned data distribution. 
            (b) For conditional reconstruction, we use MAP Gradient Ascent (MAPGA), a zero-shot posterior sampling approach that reconstructs flow fields \( \mathbf{\Psi}^{\prime}_0 \) from partial measurements \( \mathcal{S} \) using the pre-trained DDPM.
            (c) To identify optimal sensor locations, we apply the SHAP framework to attribute marginal reconstruction contributions to modular sensor subregions. The resulting attribution scores quantify the relative importance of each subregion, supporting modular, interpretable and deployable sensor selection strategies.
            }
    \end{center}
\end{figure}

Urban flows, or in broader terms turbulent flows, exhibit inherently stochastic dynamics. We leverage this stochastic nature to explore the application of diffusion models as foundational tools for sparse reconstruction and optimal sensor placement. Diffusion models are uniquely well-suited for this purpose, given their state-of-the-art ability to model complex, high-dimensional distributions directly from training data~\citep{ho2020, song2021scorebased}.

We ground the study in a high‑fidelity DNS database of flow around a wall‑mounted square cylinder, a canonical proxy for an isolated building. It is well known that the smallest turbulent scales are hardest to model for deep generative models~\citep{EIVAZI2022117038, li_generative_2023}, hence, we subtract the time-averaged mean from the velocity field $\mathbf{U}(x,y,z,t) = (U,V,W)$ from the instantaneous velocity field and study the mid‑span slice at $z/h=0$. The diffusion model is then trained and evaluated on the resulting fluctuation tensor:

\begin{equation}\label{eqn:data-fluc}
  \mathbf{\Psi}^{\prime}(x,y,t)=
  \begin{bmatrix}
    u^{\prime}(x,y,t)\\
    v^{\prime}(x,y,t)
  \end{bmatrix},
  \qquad
  u^{\prime} = U-\overline{U},\;
  v^{\prime} = V-\overline{V},
\end{equation}

\noindent
where overbars denote ensemble averages in time. Note that this preprocessing lets the DDPM focus on the dynamically rich, multi‑scale fluctuations while the mean field is considered to be known from the flow statistics. In general, the mean field $\overline{\mathbf{\Psi}}$ can be obtained from RANS or LES baseline simulations, long-time sensor averages, or historical flow climatologies, all standard resources in urban flow monitoring. For the statistically stationary flows considered in this work, the time-averaged mean is constant across all realizations; the training mean, computed from the DNS dataset, is therefore equally valid for any test snapshot and can be directly superimposed to recover the full velocity field $\mathbf{\Psi} = \overline{\mathbf{\Psi}} + \mathbf{\Psi}^{\prime}$. It is well established in the literature~\citep{corrdiff} that jointly inferring the spatially varying mean and instantaneous fluctuations from sparse single-snapshot observations is a fundamentally ill-posed problem, independent of the learning architecture employed (see Supplementary Note~7 for a formal argument). Furthermore, since the framework operates exclusively on $\mathbf{\Psi}'$, it is Galilean invariant by construction: adding any constant reference velocity $\mathbf{U}_0$ to the flow leaves $\mathbf{\Psi}' = \mathbf{\Psi} - \overline{\mathbf{\Psi}}$ unchanged, and therefore does not affect the diffusion prior, the MAPGA reconstruction, or the SHAP-based sensor placement.

The Diff--SPORT framework consists of three stages as shown in figure~\ref{fig:overview}. First, we learn the underlying probability distribution of turbulence data $p(\mathbf{\Psi}^{\prime}(x,y,t))$ by training a diffusion model as shown in figure~\ref{fig:overview}(a). Starting from Gaussian noise, the model generates high-fidelity flow samples unconditionally and further supports conditional reconstruction by maximizing the posterior \( p_\theta(\mathbf{\Psi}^{\prime}_0 | \mathcal{S}) \), where measurements are obtained as \( \mathcal{S} = H \odot \mathbf{\Psi}^{\prime}(\mathbf{x}, t) \), here \(H\) is the binary measurement mask and \( \odot \) is the Hadamard product. We solve this inverse problem using the MAP Gradient Ascent (MAPGA) algorithm ~\citep{Gutha2024InversePW} that optimizes the posterior directly through diffusion prior gradients (see, Figure~\ref{fig:overview}(b)), outperforming state-of-the-art score-based methods such as $\Pi$GDM~\citep{songPseudoinverseGuidedDiffusionModels2022} or DPS~\citep{chung2023diffusion} in terms of reconstruction accuracy and number of sensor measurements required. Finally, we integrate a Shapley-value-based~\citep{shapley1953value} sensor attribution scheme into the framework such that it computes the marginal utility of each candidate subregion via game-theory-based coalition analysis by utilizing the advanced sparse reconstruction method proposed earlier (see, Figure~\ref{fig:overview}(c)). A key advantage of our proposed methods is that they function as efficient wrappers around pre-trained diffusion models, significantly reducing computational overhead while maintaining high performance. The turbulent dynamics are ensured to remain statistically consistent over time, enabling the model to generalize and make predictions in a dynamically evolving but statistically stationary environment. Applying this framework to a simplified urban environment demonstrates that these methodological advances can improve reconstruction accuracy and computational efficiency, enabling reliable flow predictions for large‑scale urban infrastructures using optimally placed sparse observations.

\subsection*{Unconditional generation and statistical evaluation}\label{sec:results-uncond}

\begin{figure}
    \begin{center}
	    	\includegraphics[width=0.78\textwidth]{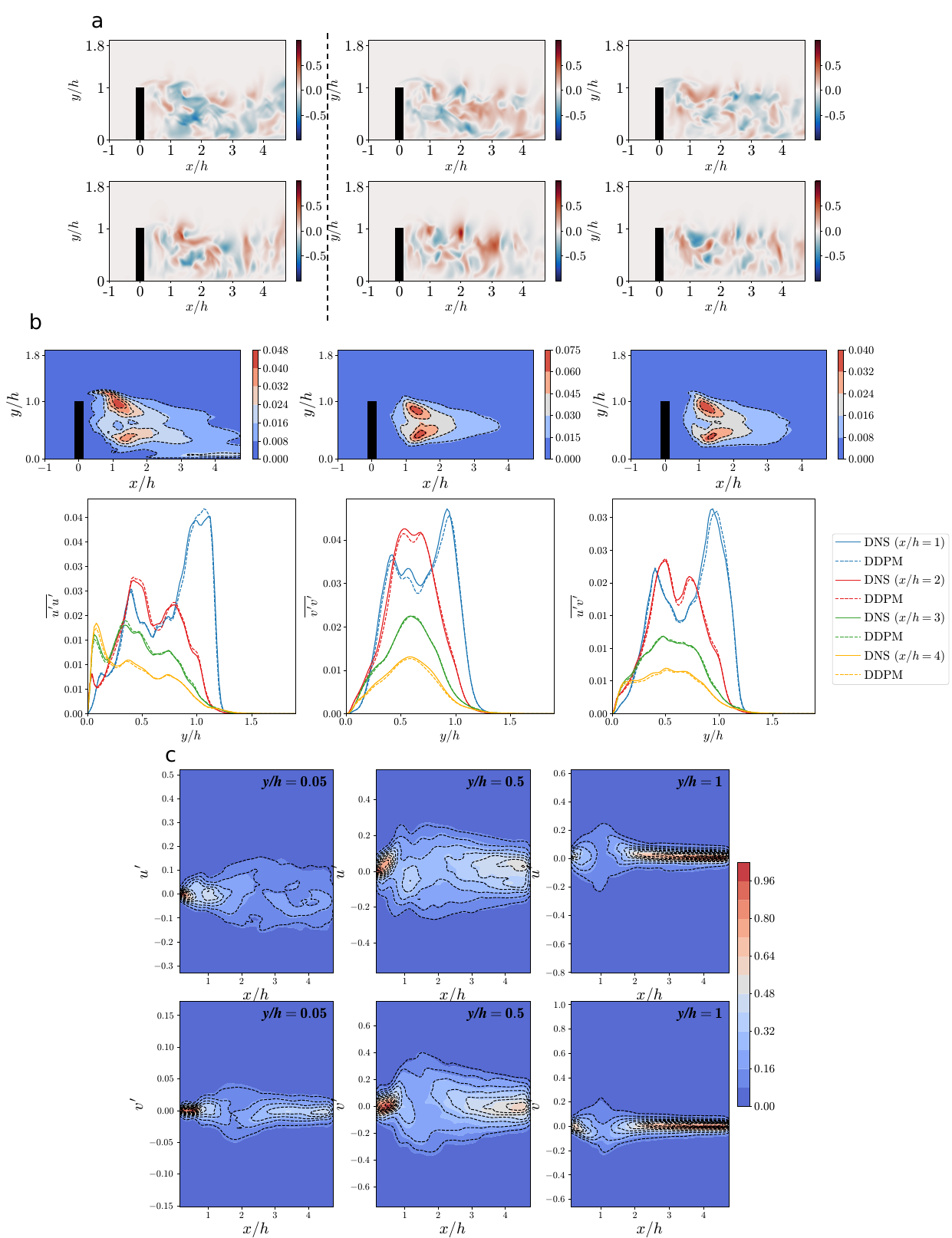}
        	  \caption{
\textbf{Evaluation of unconditional turbulent flow generation using a DDPM-based prior.} 
(a) Instantaneous velocity fluctuation fields in the (top) streamwise (\(u^{\prime}\)) and (bottom) vertical (\(v^{\prime}\)) directions, comparing DNS snapshots (left) with two unconditionally generated samples from the diffusion model (middle and right). Visual agreement demonstrates the model's ability to recover coherent structures without any conditioning. 
(b) Spatial contours and line profiles of Reynolds stress components \(\overline{u^{\prime}u^{\prime}}\), \(\overline{v^{\prime}v^{\prime}}\) and \(\overline{u^{\prime}v^{\prime}}\), averaged across the dataset. Line plots at selected streamwise locations confirm close alignment between DNS and generated statistics. 
(c) Probability density functions (PDFs) of \(u^{\prime}\) and \(v^{\prime}\) across multiple vertical positions, comparing the DNS distribution (solid contours) with that of the generated samples (dashed contours). Agreement in both large-scale distribution and fine-scale variance confirms the diffusion model's ability to approximate the underlying distribution \(p(\mathbf{\Psi}^{\prime})\) of the turbulent flow. Color scales are normalized consistently across DNS and DDPM visualizations to enable direct visual comparison. Source data are provided as a Source Data file.
}\label{fig:uncond‑eval}
    \end{center}
\end{figure}

We first evaluate the diffusion prior in an unconditional setting, i.e. without any measurement conditioning. From random Gaussian noise ($\epsilon$) the trained DDPM generates the same number of snapshots as contained in the training set, thereby sampling from an empirical approximation of the learned probability distribution $p(\mathbf{\Psi}^{\prime}(\mathbf{x, y},t))$. Figure~\ref{fig:uncond‑eval} gathers three complementary comparisons between these synthetic fields and the ground‑truth DNS data: (a) instantaneous flow visualizations, (b) second‑order turbulence statistics and (c) probability density functions (PDFs).

Figure~\ref{fig:uncond‑eval}(a) overlays randomly selected DNS and DDPM snapshots of the streamwise, $u^{\prime}$ and vertical, $v^{\prime}$, velocity fluctuations. The diffusion model recovers both the large‑scale coherent eddies and the stochastic finer scales visible in the DNS, illustrating qualitative fidelity across all the spatial locations. Figure~\ref{fig:uncond‑eval}(b) benchmarks the performance of DDPM through second-order statistics. Filled contours show Reynolds‑stress maps ($\overline{u^{\prime}u^{\prime}}$, $\overline{v^{\prime}v^{\prime}}$, $\overline{u^{\prime}v^{\prime}}$) from DNS, while dotted isolines denote the DDPM counterpart; line plots underneath extract stress profiles at four streamwise stations ($x/h {=} 1, 2, 3, 4$) as a function of the vertical coordinate $y/h$. Synthetic and reference exhibit an excellent level of agreement, confirming that the generative prior accurately reproduces second-order turbulent statistics. This includes the mean-square velocity fluctuations, $\overline{u^{\prime}u^{\prime}}$ and $\overline{v^{\prime}v^{\prime}}$, which quantify the turbulent kinetic energy in the streamwise and vertical directions, respectively. Crucially, our method also captures the Reynolds shear stress component $\overline{u^{\prime}v^{\prime}}$, which governs the vertical transport of streamwise momentum, an essential mechanism in wall-bounded turbulence. Figure~\ref{fig:uncond‑eval}(c) presents PDFs of $u^{\prime}$ and $v^{\prime}$ as a function of the non-dimensional streamwise coordinate $x/h$ at three vertical locations $y/h {=} 0.05, 0.5, 1$. The filled contours represent DNS, while dotted contours denote DDPM velocity fluctuations. The PDF comparisons span a broad range of vertical locations, confirming that the generative prior faithfully captures the changing structure of the turbulent fluctuations across the domain. Also, the excellent overlap across all heights shows that the model captures not only first‑ and second‑order moments but also the full bivariate distribution of turbulent fluctuations.

In summary, Fig.~\ref{fig:uncond‑eval} demonstrates that the DDPM acts as a high-fidelity generative surrogate for DNS. It captures both the structure and statistics of turbulence without conditioning, generating samples that are visually realistic and statistically consistent, which is a critical prerequisite for the conditional reconstructions that follow.

\subsection*{Sparse reconstruction through conditional generation }\label{sec:results-cond}

\begin{figure}
    \begin{center}
	    	\includegraphics[width=1\textwidth]{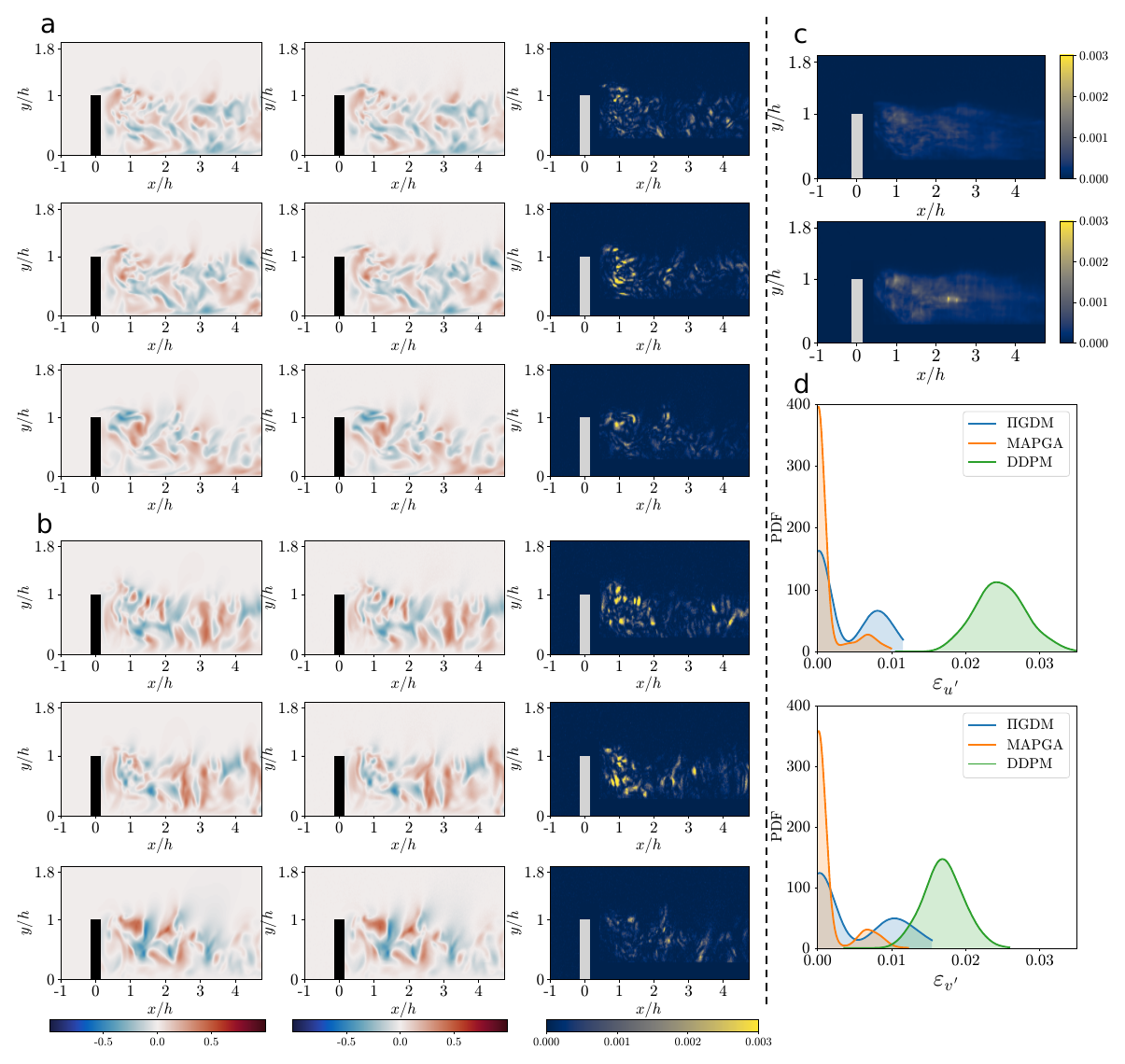}
        	\caption{
\textbf{Evaluation of conditional turbulent-flow reconstruction using MAP Gradient Ascent (MAPGA) and comparison with \( \Pi \)GDM and unconditional generation (DDPM).} Instantaneous reconstruction of velocity fluctuation fields in the (a) streamwise (\( u^{\prime} \)) and (b) vertical (\( v^{\prime} \)) directions for three representative snapshots (each represented in a different row). Here we show the DNS (left), MAPGA reconstruction (middle) and the corresponding squared error field (\(\varepsilon\), right). Despite no sensor availability in the wake region, MAPGA successfully recovers both large-scale and fine-scale flow features.
(c) Mean squared error fields for \( u^{\prime} \) (top) and \( v^{\prime} \) (bottom), temporally averaged over the test set ($n = 1000$). Error concentrations appear in wake and shear regions, but overall error magnitudes remain low due to the robustness of MAPGA reconstructions.
(d) Probability density functions (PDFs) of spatially averaged reconstruction error \( \varepsilon \) for \( u^{\prime} \) (top) and \( v^{\prime} \) (bottom), comparing MAPGA, \( \Pi \)GDM and unconditional generation via DDPM ($n = 1200$). MAPGA consistently exhibits the lowest error and sharpest peak, confirming its effectiveness for zero-shot inference. Source data are provided as a Source Data file.
}\label{fig:cond‑eval}
    \end{center}
\end{figure}

Having established the diffusion model as a reliable prior for turbulent flow synthesis, we next perform conditional generation using masked sensor inputs. One of the major motivations behind this work is to develop a practical and effective methodology for sparse reconstruction of turbulent flows from limited sensor observations. To this end, we design a baseline mask (see, Methods section) that retains approximately 15\% of the available flow field information, specifically the non-wake regions with low energy content, where sensor placements are realistically feasible.

As discussed above, conditional generation reconstructs turbulent flow fields by integrating the learned probabilistic prior \(p_{\theta}(\mathbf{\Psi}^{\prime}(\mathbf{x}, t))\) with sparse sensor measurements ($\mathcal{S}$). The effectiveness of conditional generation methods hinges upon accurately reconstructing the underlying flow field \(\mathbf{\Psi}^{\prime}_0\) by maximizing the posterior probability \(p_{\theta}(\mathbf{\Psi}^{\prime}_0|\mathcal{S})\). By Bayes' theorem, this posterior is proportional to the product of the prior \(p_{\theta}(\mathbf{\Psi}^{\prime})\) and the likelihood \(p(\mathcal{S}|\mathbf{\Psi}^{\prime})\). Numerous Bayesian inference techniques have been developed recently to solve such conditional reconstruction problems, including methods such as DPS~\citep{chung2023diffusion}, DDRM~\citep{kawarDenoisingDiffusionRestoration2022} and $\Pi$GDM ~\citep{songPseudoinverseGuidedDiffusionModels2022}. In this work, we utilize another Bayesian approach, maximum a posteriori (MAP)-based gradient estimation~\citep{Gutha2024InversePW}, designed to leverage gradient-based optimization informed directly by the learned diffusion prior. We benchmark the performance of MAPGA against the current state-of-the-art, the $\Pi$GDM method. The $\Pi$GDM approach addresses inverse problems by estimating conditional scores derived directly from the measurement model, enabling effective handling of noisy, nonlinear, or even non-differentiable measurements without additional training. In contrast, our MAPGA framework explicitly formulates inverse problems as MAP estimation tasks. By directly optimizing the posterior probability through gradient-based methods, leveraging the learned diffusion prior, MAPGA circumvents the approximations inherent in score-based techniques like $\Pi$GDM. This direct gradient optimization allows MAPGA to more effectively integrate measurement information, yielding improved reconstruction accuracy, particularly in challenging sparse-measurement scenarios.

For quantitative evaluation, we primarily compute the pixel-wise absolute error field (\(\varepsilon\)) relative to the ground truth, as well as the mean squared error (MSE), defined as:

\begin{equation}
    \varepsilon_{MSE} = \frac{1}{N_{x}}\frac{1}{N_{t}}\sum_{l=1}^{N_t}\sum_{k=1}^{N_x}(u^{\prime}_{kl,\text{DNS}} - u^{\prime}_{kl,\text{Pred}})^2,
\end{equation}
 
\noindent

where \(u^{\prime}_{kl,\text{DNS}}\) and \(u^{\prime}_{kl,\text{Pred}}\) represent the DNS and predicted velocity fluctuations at spatial location \(k\) and timestep \(l\), respectively, evaluated independently in both the streamwise (\(u^{\prime}\)) and vertical (\(v^{\prime}\)) directions, while $N_x$ and $N_t$ are the number of grid points in the field and timesteps. The average error across both components is then computed as \( 
\overline{\varepsilon} = 0.5 (\varepsilon_{u^{\prime}} + \varepsilon_{v^{\prime}})\).

Figure~\ref{fig:cond‑eval} provides a comprehensive evaluation of the MAPGA method for reconstructing turbulent velocity fields from sparse sensor measurements, using the baseline mask configuration shown in Figure~\ref{fig:overview}(c). Figures~\ref{fig:cond‑eval} (a) and (b) present qualitative comparisons of three representative time snapshots for the streamwise ($u^{\prime}$) and vertical ($v^{\prime}$ ) components, respectively. Each row displays the ground-truth DNS field (left), the MAPGA reconstruction (middle) and the corresponding error (right). These visualizations demonstrate that MAPGA successfully recovers both large-scale coherent structures and fine-scale turbulent fluctuations, yielding instantaneous reconstructions with DNS-level fidelity despite limited measurements from non-wake regions. Figure~\ref{fig:cond‑eval}(c) shows the mean squared error fields for $u^{\prime}$ and $v^{\prime}$, temporally averaged across the test set. These maps show that the reconstruction errors are primarily localized within the wake region, but their overall magnitude remains low, demonstrating MAPGA’s ability to accurately recover wake dynamics using only near-wall measurements. Notably, the mean error magnitude is higher for \( v^{\prime}\) than for \( u^{\prime}\), likely due to stronger relative fluctuations in the vertical direction. Figure~\ref{fig:cond‑eval}(d) presents probability density functions (PDFs) of the spatially averaged reconstruction error $\varepsilon$ for each method across the entire test dataset. MAPGA achieves a sharp distribution peak around $\mathcal{O}(\overline{\varepsilon}) = 10^{-3}$ indicating both high fidelity and low variance. In contrast, \(\Pi\)GDM produces a diffused error distribution with multiple peaks, indicating less stable reconstructions. While unconditional DDPM sampling performs significantly worse than the sparse reconstruction methods in absolute terms, it is important to note that it already captures much of the underlying flow structure, effectively performing the bulk of the generative task without any conditioning. Quantitatively, MAPGA outperforms $\Pi$GDM by nearly a factor of three in terms of accuracy and precision. For completeness, we also benchmark MAPGA against Diffusion Posterior Sampling (DPS)~\citep{chung2023diffusion}, which differs from $\Pi$GDM only by a scaling factor in its formulation. As shown in Supplementary Note~1 (Supplementary Fig.~1), MAPGA maintains its performance advantage over DPS with comparable error reduction, confirming that the improvements stem from MAPGA's direct posterior optimization rather than limitations specific to $\Pi$GDM. Together, these results highlight MAPGA's precision, stability and suitability for real-world deployment in sparse-sensing regimes. To further substantiate practical utility, Supplementary Note~3 reports rigorous robustness and generalization checks: reconstruction performance under varying levels of measurement noise (SNR~10--25~dB), robustness to sensor misregistration (spatial shifts and rotations), and performance on out-of-distribution spanwise planes unseen during training. Across all three scenarios, MAPGA degrades gracefully without catastrophic failure, confirming its suitability for realistic deployment conditions.

\subsection*{Optimal sensor placement}\label{sec:results-osp}

\begin{figure}
    \begin{center}
	    	\includegraphics[width=1\textwidth]{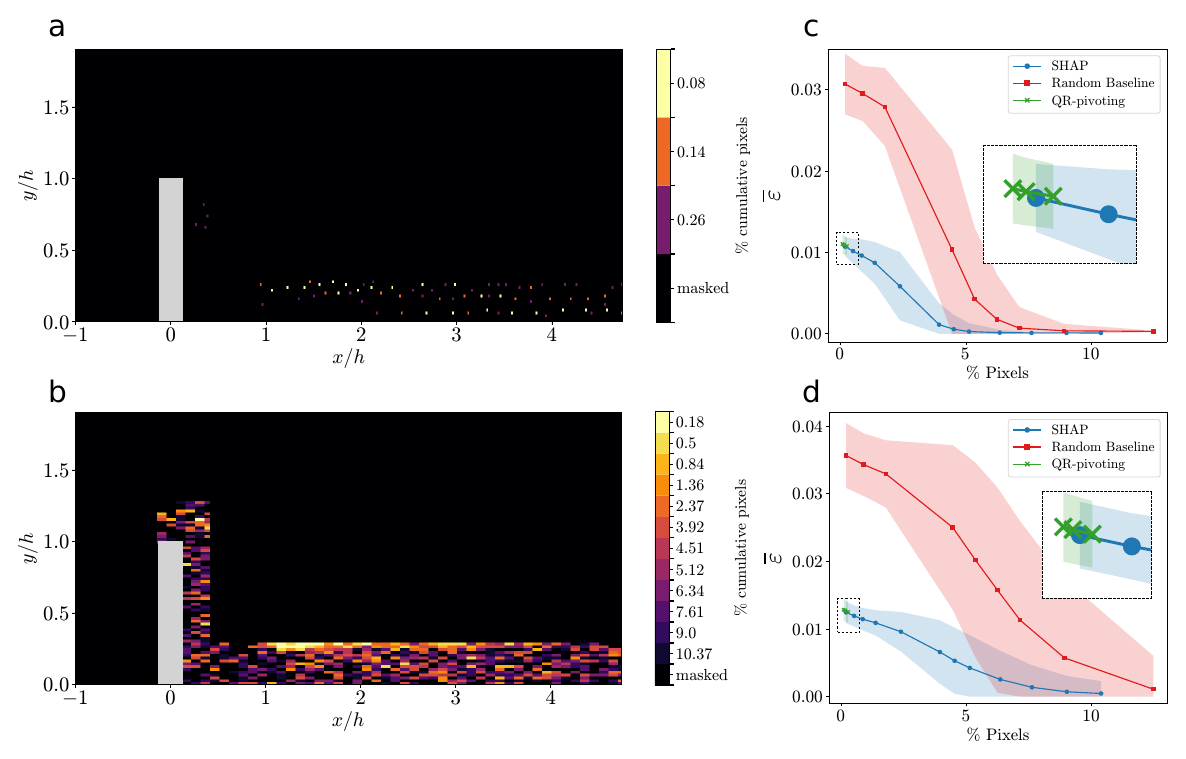}
\caption{\label{fig:shap-osp}
\textbf{Comparison of sensor attribution and reconstruction performance using SHAP, QR-pivoting and random placement.}
(a) Sensor importance map computed via QR-pivoting on training snapshots, restricted to the feasible baseline region. The selected sensor locations are spatially scattered and disconnected, limiting deployability. 
(b) SHAP-based importance map, obtained by averaging marginal contributions across the training set. Thresholding yields spatially coherent sensor configurations at varying sparsity levels. High-importance regions from SHAP and QR-pivoting largely overlap, although SHAP supports finer control over sensor budgets. 
(c)-(d) Mean reconstruction error \( \overline{\varepsilon} \) for \( u' \) and \( v' \) as a function of sensor coverage, comparing SHAP, QR-pivoting and random placement. (c) Reports the error evolution after selecting the lowest MAPGA error per test field by isolating data-driven variability. Centre lines denote the mean and shaded bands $\pm$ one standard deviation across $n = 1000$. (d) Mean error and uncertainty over multiple MAPGA runs per field, capturing both test-set and MAPGA inference variability. Centre lines denote the mean and shaded bands $\pm$ one standard deviation across $n = 1000$ $\times$ six MAPGA runs. SHAP consistently outperforms random placement. The figure shows SHAP being slightly better QR-pivoting in low-sensor configurations (see insets). Unlike QR, SHAP is not limited by the number of training snapshots and allows flexible scaling of sensor counts. Source data are provided as a Source Data file.
}
    \end{center}
\end{figure}

While the previous section demonstrates that MAPGA can reconstruct turbulent flow fields with high accuracy using the baseline mask, the performance of such reconstruction also depends on the spatial configuration of the sensors. However, in real-world scenarios,
sensor deployment is further constrained by cost along with accessibility and physical limitations, which raises a fundamental question: which sensor locations contribute most to accurate flow reconstruction? To address this, we introduce a data-driven optimal sensor placement framework based on SHAP values. By quantifying the marginal contribution of each candidate sensor region to the overall reconstruction accuracy using coalitions based on cooperative game theory, the SHAP framework enables us to rank sensor locations by importance.

We evaluate the Shapley-based ranking against two baselines: (i) random selection and (ii) QR pivoting~\citep{Manohar2017DataDrivenSS}. A comprehensive description of baseline methods is presented in the Methods section.

Figure~\ref{fig:shap-osp} compares SHAP-based sensor placement with QR-pivoting and a random baseline, focusing on both spatial attribution and reconstruction accuracy. Figure~\ref{fig:shap-osp}(a) shows QR-derived sensor importance based on pivoted QR decomposition applied to the training set. As mentioned in the methods section, for all three OSP strategies, we only select sensors from the feasible baseline region. It must be noted that the selections within this region for QR are spatially scattered and disconnected, which complicates real-world deployment. In contrast, Figure~\ref{fig:shap-osp}(b) shows the SHAP-based sensor importance map, which reflects averaged importance scores across the training set. Thresholding the resulting SHAP values yields spatially coherent sensor configurations for multiple sparsity levels. Importantly, the regions of high utility identified by QR-pivoting and SHAP follow a similar trend in low sensor regime, whilst SHAP is capable of providing more sensor locations as per the need. 

Figures~\ref{fig:shap-osp}(c) and (d) plot the average reconstruction error \( \overline{\varepsilon} \) for \( u' \) and \( v' \), respectively, as a function of sensor coverage (\% active pixels). All reconstructions are performed using MAPGA with fixed masks for QR-pivoting and SHAP and seven different masks for the random baseline for each sensor count. Each sensor configuration is evaluated on the full test set. In panel~(c), for every test field and MAPGA run, we select the best (i.e., lowest-error) reconstruction among multiple MAPGA runs. Isolating the variability due to data and suppressing probabilistic uncertainty from the inference process. The best selection strategy can be utilized in practical scenarios, leveraging the temporal error evolution curves for best results. In contrast, panel~(d) reports the average error and standard deviation across all MAPGA runs per field, thus incorporating both stochastic reconstruction uncertainty and test-set variability. This offers a more realistic measure of expected reconstruction fidelity in practical deployment settings. Across both evaluations, SHAP-selected sensors outperform the random baseline and match or exceed the performance of QR, particularly under tight sensor budgets (see insets). Unlike QR, which is fundamentally limited by the number of training snapshots, SHAP enables flexible scaling. A comprehensive sensitivity analysis of the SHAP attribution with respect to coalition size and segmentation granularity, together with a full computational cost breakdown of the Diff--SPORT pipeline, is provided in Supplementary Notes~4 and~5.

\subsection*{Performance on experimental scalar transport in a realistic multi-building urban canopy}
\label{sec:results-exp}

\begin{figure}
    \begin{center}
        \includegraphics[width=1\textwidth]{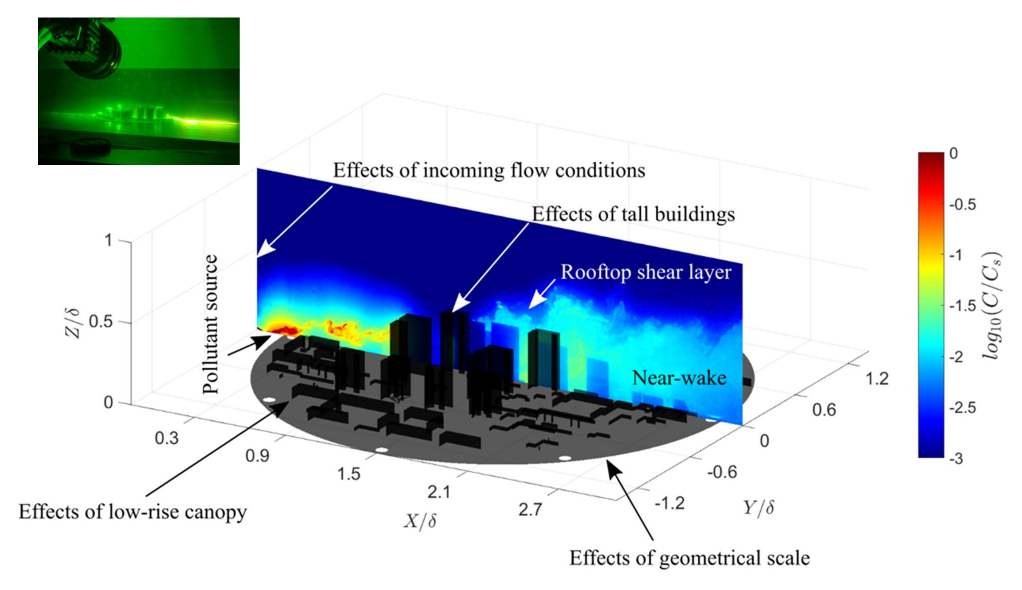}
        \caption{\label{fig:exp-setup}
        \textbf{Experimental setup and urban canopy geometry.}
        1:2400 scale water-flume model of the Beijing Haidian neighbourhood (laboratory photograph,
        inset) and representative instantaneous concentration field $C/C_S$,
        annotating the dominant physical mechanisms that shape scalar dispersion: approaching boundary
        layer, rooftop shear layer over the tall cluster, near-wake recirculation and the modulating influence of the low-rise canopy. Adapted from Ref.~\cite{PLIF_ref}, distributed under the Creative Commons Attribution 4.0 International License (\nolinkurl{https://creativecommons.org/licenses/by/4.0/}).
        }
    \end{center}
\end{figure}

The preceding sections establish Diff--SPORT's capabilities in a controlled, single-obstacle DNS setting. To assess whether the framework transfers to fundamentally different conditions, we implement it, without any algorithmic modifications to the LT2400 experimental PLIF dataset \citep{PLIF_ref} detailed in the Methods section. The three framework components (DDPM prior, MAPGA, and SHAP) are reused directly, only the diffusion prior is retrained on the concentration data. This constitutes a strict three-axis generalization test: (i)~from velocity fluctuations to passive scalar concentration, (ii)~from $\mathrm{Re} = 2000$ to $\mathrm{Re} \approx 14{,}000$, and (iii)~from a single square obstacle to a 14-building tall cluster embedded in a multi-scale low-rise urban canopy. Figure~\ref{fig:exp-setup} illustrates the experimental geometry and a representative concentration field $C/C_S$, annotating the dominant physical mechanisms that shape scalar dispersion: the approaching boundary layer, the rooftop shear layer over the tall cluster, near-wake recirculation, and the modulating influence of the low-rise canopy \citep{PLIF_ref}. These features introduce substantially more complex spatial structure, making LT2400 a demanding benchmark for generative reconstruction.

\begin{figure}
    \begin{center}
        \includegraphics[width=0.67\textwidth]{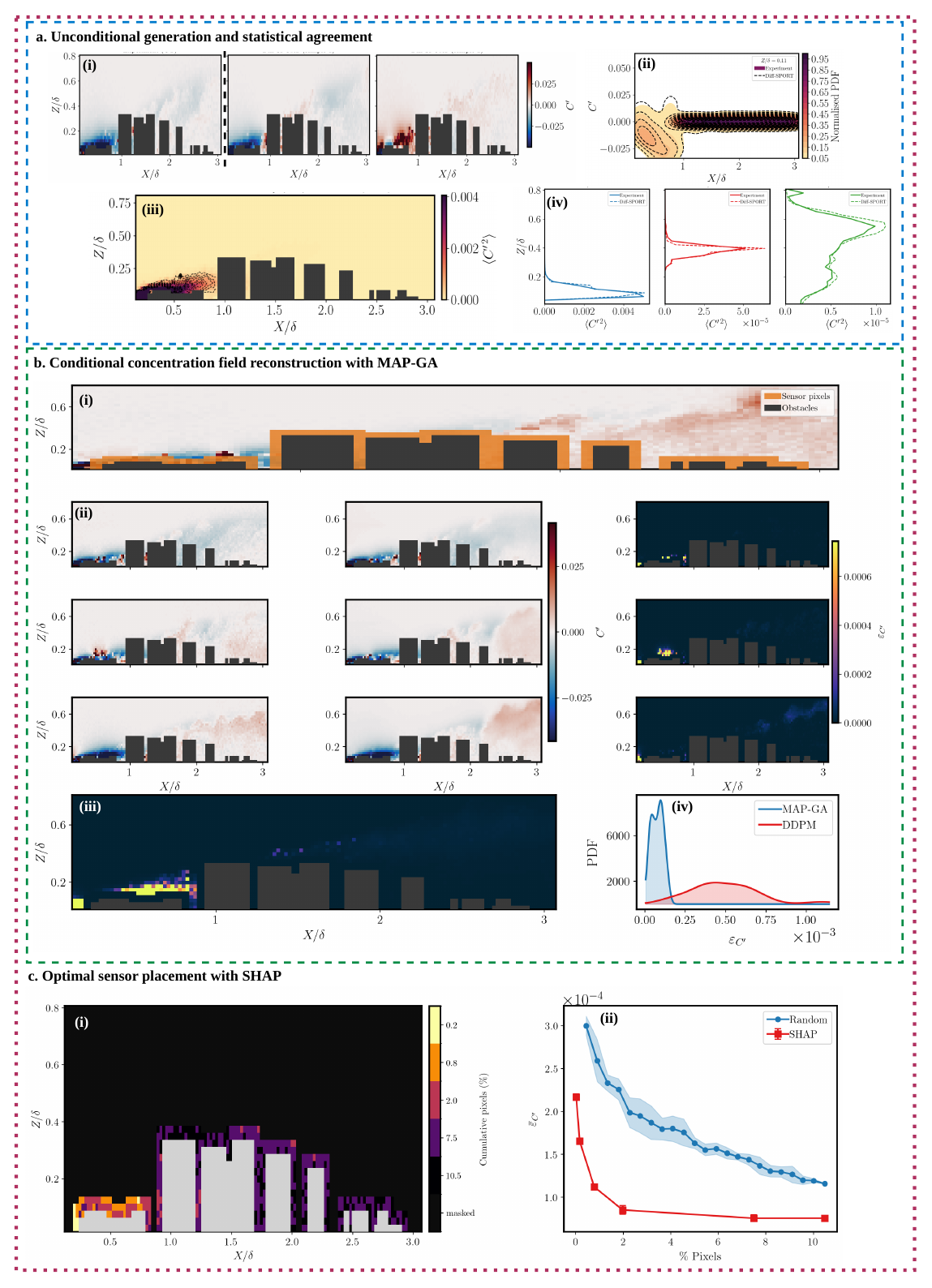}
        \caption{\label{fig:exp-eval}
        \textbf{Application of Diff--SPORT to experimental PLIF scalar transport data from
        the LT2400 urban canopy.}
        (a) Unconditional generation and statistical agreement: (i) representative
        ground-truth PLIF and Diff--SPORT-generated concentration fluctuation snapshots ($C^{\prime}$); (ii)
        streamwise-evolving PDF of $C^{\prime}$ at $Z/\delta = 0.11$, comparing
        Experiment (filled contours) and Diff--SPORT (dashed contours); (iii)
        concentration variance $\langle C^{\prime 2}\rangle$ map: Experiment (filled
        contours) vs Diff--SPORT (dashed); (iv) concentration variance profiles $\langle C^{\prime
        2}\rangle$ at $X/\delta = 0.49$, $1.49$ and $2.49$.
        (b) Conditional concentration field reconstruction with MAPGA:
        (i) feasibility mask applied to a representative snapshot, retaining 429 sensor
        pixels (${\approx}10.5\%$ of the $128{\times}32$ domain) near ground level and
        building facades;
        (ii) three randomly selected test snapshots each displaying ground-truth
        concentration (left), MAPGA reconstruction (centre) and instantaneous squared
        error ($\varepsilon_{C^{\prime}}$, right);
        (iii) mean squared error field ($\varepsilon_{C^{\prime}}$) averaged over the full test dataset;
        (iv) error PDFs comparing MAPGA and DDPM ($n = 65$).
        (c) Optimal sensor placement with SHAP: (i) SHAP importance map over the canopy
        domain with building footprints masked; (ii) mean reconstruction error
        $\bar{\varepsilon}$ as a function of sensor coverage, comparing SHAP-selected and
        randomly placed configurations. Centre lines denote the mean and shaded bands $\pm$ one standard deviation across random masks ($n = 195$); SHAP error bars denote $\pm$ one standard deviation across four MAPGA seeds ($n = 260$). Source data are provided as a Source Data file.
        }
    \end{center}
\end{figure}

We first evaluate the diffusion prior trained on LT2400 in an unconditional setting. Figure~\ref{fig:exp-eval}(a) presents three complementary comparisons between DDPM-generated and ground-truth PLIF concentration fields. Sub-panel~(i) overlays representative PLIF snapshots with unconditionally generated samples. The model recovers the large-scale plume
structure emanating from the pollutant source, the attenuation of scalar fluctuations within
the low-rise canopy, and the elevated $C^{\prime}$ regions associated with rooftop ejections above the tall cluster, all without any conditioning. Sub-panel~(ii) presents the streamwise-evolving PDF of $C^{\prime}$ at $Z/\delta = 0.11$;
the Diff--SPORT contours closely follow the experimental distribution, capturing both the
broad bimodal spread near the source region ($X/\delta \approx 0.5$--$1.0$) and the
progressive narrowing downstream. Sub-panels~(iii) and~(iv) benchmark second-order statistics
through the concentration variance map $\langle C^{\prime 2}\rangle$ and vertical profiles
at $X/\delta = 0.49$, $1.49$ and $2.49$, Diff--SPORT reproduces the rapid variance attenuation
from $\mathcal{O}(10^{-3})$ upstream to $\mathcal{O}(10^{-5})$ within and downstream of the tall cluster, with close quantitative agreement at all three stations. Together, these results establish the pretrained
diffusion prior as a reliable generative surrogate for scalar transport in a complex urban
canopy, a prerequisite for the conditional reconstructions that follow.

Having established the diffusion prior on LT2400, we next perform conditional
reconstruction of the concentration field using MAPGA. The feasibility mask retains 429
sensor pixels (${\approx}10.5\%$ of the $128{\times}32$ domain), restricted to near-ground
regions and building facade surfaces while excluding the wake of the tall cluster, where
sensor deployment is physically impractical. Figure~\ref{fig:exp-eval}(b)(i) illustrates
this mask on a representative snapshot. Sub-panel~(ii) shows MAPGA reconstructions for
three randomly selected test snapshots, across test cases, MAPGA recovers the spatial extent of the scalar plume, the
attenuation pattern through the low-rise canopy, and the concentration gradient at the
rooftop shear layer, using only the sparse near-wall observations. Sub-panel~(iii) presents
the mean absolute error field over the test dataset, errors are largest in the near-wake of
the tall cluster, consistent with the high-gradient, sensor-deprived region, yet their
overall magnitude remains low. Sub-panel~(iv) shows that MAPGA yields a sharper, lower-error
distribution, confirming that zero-shot posterior
sampling transfers effectively to experimental scalar data in a complex multi-building
environment.

Finally, we apply the SHAP-based sensor placement framework to the LT2400
domain. Figure~\ref{fig:exp-eval}(c)(i) shows the SHAP importance map over the feasible
canopy region, the highest-ranked sensors (${\lesssim}0.8\%$ cumulative pixels) are
concentrated at ground level upstream of the tall cluster ($X/\delta \approx 0.3$--$0.7$)
and within the gaps between tall building facades, locations that intercept the incoming
scalar plume before it interacts with the cluster, reflecting a physically interpretable
placement strategy. We note that the wake of the tall building cluster falls outside the
domain-informed feasibility mask (Methods) by design; within the feasible region, SHAP's
upstream preference reflects the advection-dominated scalar transport regime rather than a
general undervaluing of high-turbulence locations, a distinction verified directly using a
velocity dataset from the same facility (Supplementary Note~8). Figure~\ref{fig:exp-eval}(c)(ii) compares reconstruction error against
random placement as a function of sensor coverage. SHAP outperforms random across the entire range, with the largest gain at ${\approx}2\%$ pixel coverage where SHAP achieves
${\approx}57\%$ lower error than random. Critically, SHAP plateaus at this $2\%$ threshold, recovering near-optimal reconstruction accuracy with only a fraction of the available sensors, while random placement requires the full $10.5\%$ to approach a comparable, yet still
${\approx}37\%$ higher error level. These results demonstrate that SHAP-guided sensor
placement transfers directly to experimental urban canopy data, identifying compact,
physically meaningful sensor configurations for scalar concentration monitoring.

\section*{Discussion}\label{sec:Discussion}

The Diff--SPORT framework advances the frontier of turbulence modeling by integrating deep generative modeling, sparse reconstruction and optimal sensor placement within a single, modular pipeline.  In contrast to prior studies that typically focus on these aspects, Diff--SPORT enables zero-shot probabilistic generation, high-fidelity reconstruction and interpretable sensor-location ranking, all using a single diffusion prior working as a foundation model for downstream tasks, offering a highly reusable prior for downstream inference problems without retraining. A preliminary assessment of the framework's robustness under geometric distribution shift, i.e. applying the unretrained LT2400 prior to an unseen canopy geometry, is also studied in Supplementary Note 9.

The MAPGA algorithm for sparse reconstruction proposed in this work significantly outperforms state-of-the-art score-based methods $\Pi$GDM and DPS, achieving lower reconstruction errors across both instantaneous flow fields and statistical quantities with fewer sensors. Importantly, MAPGA achieves this accuracy without utilizing any sensor measurements from the wake region, a domain typically favored by conventional methods for its high-energy content. However, placing sensors in the wake region is often impractical due to physical obstructions and flow disturbances. Diff--SPORT addresses this challenge by incorporating a domain-informed mask design that enforces realistic sensor feasibility constraints. Furthermore, the use of Shapley-based sensor attribution enables the identification of compact, spatially contiguous subregions, offering a marked improvement in deployability compared to the scattered pixel placements produced by linear methods such as QR pivoting. This modular approach to sensor selection allows Diff--SPORT to flexibly adapt to various practical constraints, making it suitable for diverse real-world deployment scenarios.

Despite the cost of DNS data generation using \texttt{Nek5000}, downstream tasks like reconstruction and sensor optimization are computationally efficient, relying on pre-trained models. Both unconditional and conditional samples are generated via independent forward passes, allowing scalable, parallel inference. In our setup, training on one NVIDIA A100 GPU took ~24 hours, while inference required ~12 seconds per snapshot. Batched generation and parallel sensor evaluation further reduce wall-clock time, supporting near real-time deployment.

A key architectural decision in Diff--SPORT is the use of an IID snapshot prior rather than a spatiotemporal prior. This is not a limitation but a deliberate choice critical for practical deployment, in real-world urban sensing, observations are inherently asynchronous, irregularly sampled, and prone to gaps, conditions under which spatiotemporal priors, which require temporally synchronized, regularly sampled measurement streams, are fundamentally ill-suited. The IID formulation conditions on any instantaneous sparse observation without temporal constraints, making it directly compatible with heterogeneous urban sensor networks at scale. This design does not sacrifice temporal statistics: as shown in Supplementary Note~6, samples from the trained prior reproduce the power spectral density of DNS velocity fluctuations across all wall-normal locations, confirming that the stationary ergodicity assumption is well-founded. Spatiotemporal priors remain a complementary approach for applications requiring explicit trajectory generation or long-horizon forecasting, a distinct use case with fundamentally different data and deployment requirements.

The generalizability of Diff--SPORT as a framework is validated across multiple axes. Out-of-distribution robustness under unseen spanwise planes is reported in Supplementary Note~3, and applicability to a different physical quantity (scalar concentration), a higher Reynolds number, and a complex multi-building configuration is directly demonstrated via the LT2400 experimental dataset (Results section), where the MAPGA reconstruction and SHAP-based sensor placement are reused directly, without algorithmic modification, on top of a diffusion prior retrained for the concentration field at $\mathrm{Re} \approx 14{,}000$ in a 14-building urban canopy.

Several avenues remain for further development. Generalization can also be enhanced by training the DDPM across different flow conditions by training foundation models. Additionally, utilizing an end-to-end differentiable sparse reconstruction technique could enable the use of differentiable attribution methods and lower the computational cost of sensor optimisation, enabling real-time deployment at scale.

Diff--SPORT’s probabilistic backbone enables it to capture the inherently stochastic nature of turbulent flows. The diffusion prior models complex, multimodal distributions in high-dimensional space, offering a natural way to express uncertainty in flow reconstructions.
Its probabilistic design, modular structure and strong empirical performance suggest promising applications not only in wind sensing and microclimate mapping but also in pollutant tracking, energy-efficient ventilation planning and data-driven environmental control in smart cities. By enabling high-fidelity reconstructions with minimal sensing infrastructure, Diff--SPORT supports scalable, real-time monitoring frameworks that align with the goals of sustainable and resilient urban development.

\section*{Methods}\label{sec:Methods}
\subsection*{Numerical simulation and flow description}\label{sec:numericalsetup}

The dataset, documented in Ref.~\cite{MartinezSanchez2023}, was generated via DNS of incompressible flow around a wall-mounted square obstacle, using the open-source spectral element code \texttt{Nek5000}\citep{fischer2008nek5000}. The obstacle has a width‑to‑height ratio of $b/h=0.25$, while the Reynolds number based on $h$ and free-stream velocity $U_{\infty}$ is $\mathrm{Re}_h = 2000$. The computational domain spans \( -1 \leq x/h \leq 5 \), \( 0 \leq y/h \leq 2 \) and \( -1.5 \leq z/h \leq 1.5 \), with spectral resolution totaling approximately 21.8 million grid points. For compatibility with machine learning models, the velocity field is spectrally interpolated onto a uniform mesh with resolution \( (N_x, N_y, N_z) = (300, 100, 150) \). 

Temporal discretization is expressed in convective time units, with a timestep \( \Delta t_s = 0.005 \), ensuring accurate resolution of unsteady turbulent features. After removing initial transients, the dataset contains 26{,}000 fully developed snapshots, covering 130 convective time units. The data is split into a 95\%-5\% for training and testing, with temporal ordering preserved to simulate realistic prediction settings. For the present study, we extract a two-dimensional slice at \( z/h = 0 \), capturing streamwise (\(x\)) and vertical (\(y\)) variations in the velocity field. Further details on the simulation setup can be found in Ref.~\cite{MartinezSanchez2023} and additional validations of this method in turbulent flows in Ref.~\cite{vinuesa2015minimum}.

\subsection*{Water-flume experiment and PLIF scalar dataset (LT2400)}\label{sec:expsetup}

The LT2400 dataset is obtained from planar laser-induced fluorescence (PLIF) measurements conducted in a closed-loop water flume~\citep{PLIF_ref}. The experimental geometry represents
a 1:2400 scale replica of the Beijing Haidian neighbourhood, comprising a 14-building tall
cluster embedded within a low-rise urban canopy. Low-rise building heights span 2, 4, and
6\,mm in model scale (4.8, 9.6, and 14.4\,m full scale), while the tall cluster ranges from
20 to 30\,mm (48--72\,m full scale). The flume test section is 6750\,mm long and 1200\,mm
wide, with a water depth of 600\,$\pm$\,1\,mm. Flow conditions are characterised by a
free-stream velocity $U_{\infty} = 0.46$\,m/s, a boundary-layer thickness $\delta_{\infty} =
83$\,mm, and a Reynolds number $\mathrm{Re} \approx 14{,}000$ based on the height of the tallest
building.

Passive scalar transport is traced using a neutrally buoyant Rhodamine 6G solution at source
concentration $C_S = 0.3$\,mg/L and Schmidt number $Sc = 2500 \pm 300$. High-resolution PLIF
snapshots are downsampled to $128 \times 32$ pixels to reduce computational cost while
retaining the dominant spatial structure of the scalar plume. Consistent with the preprocessing
in equation~(\ref{eqn:data-fluc}), the time-averaged mean concentration $\overline{C}$ is
subtracted prior to training, and the DDPM is trained on the resulting scalar fluctuations
$C^{\prime} = C - \overline{C}$. The training set is augmented by generating multiple views per raw snapshot through offset-based spatial downsampling in $x$ direction, increasing the number of training samples from the limited experimental acquisition.
Unlike the DNS velocity fluctuations, scalar concentration fluctuations exhibit
heavy-tailed distributions and high intermittency; we therefore apply a
sign-preserving logarithmic normalization
$\tilde{C}^{\prime} = \mathrm{sign}(C^{\prime})\,\log(1 + |C^{\prime}|/\alpha)/\beta$
($\alpha = 1.0$,\, $\beta = 7.0$) prior to training, compressing large-amplitude
events to the $[-1,1]$ range required by the DDPM. All three framework components, the DDPM prior, MAPGA
reconstruction, and SHAP-based sensor placement, are applied without algorithmic modification,
only the diffusion prior is retrained on the concentration data.

\subsection*{Diffusion models for unconditional data generation} \label{sec:method-cond}

The foundation of our sparse reconstruction approach is to leverage a strong, data-driven prior in the form of a generative model. Following the success of diffusion models in computer vision, we use a denoising diffusion probabilistic model (DDPMs)~\citep{ho2020} as a generative priors for turbulent flows. Gaussian noise is iteratively added to a sample $\mathbf{\Psi}^{\prime}_{0}(\mathbf{x}, t)$ in the forward process $\mathbf{\Psi}^{\prime}_0(\mathbf{x}, t) \to  \mathbf{\Psi}^{\prime}_{1}(\mathbf{x}, t) \to \dots \to  \mathbf{\Psi}^{\prime}_{T}(\mathbf{x}, t)$, where the subscripts index the diffusion time $\tau \in  [0,T]$. The model is tasked with learning the reverse process $ \mathbf{\Psi}^{\prime}_{T}(\mathbf{x}, t) \to \dots \to \mathbf{\Psi}^{\prime}_{1}(\mathbf{x}, t) \to \mathbf{\Psi}^{\prime}_{0}(\mathbf{x}, t)$ that denoises $ \mathbf{\Psi}^{\prime}_{\tau}(\mathbf{x}, t)$ iteratively. For a particular choice of transition kernel in the forward process, see equation~(\ref{eqn:ddpmfwdkernel}), it follows that $p( \mathbf{\Psi}^{\prime}_{T}(\mathbf{x}, t))$ is a standard Gaussian distribution. Consequently, passing the noisy samples $\mathbf{\Psi}^{\prime}_{T}(\mathbf{x}, t) \sim \mathcal{N}(\bm 0,\bm I)$ through the learned reverse process creates clean data samples $\mathbf{\Psi}^{\prime}_{0}(\mathbf{x}, t)$.  
\begin{subequations}
    \begin{equation}\label{eqn:ddpmfwd}
        q( \mathbf{\Psi}^{\prime}_{1:T}(\mathbf{x}, t) | \mathbf{\Psi}^{\prime}_{0}(\mathbf{x}, t)) = \textstyle\prod_{\tau=1}^T q(\mathbf{\Psi}^{\prime}_{\tau}(\mathbf{x}, t)| \mathbf{\Psi}^{\prime}_{\tau-1}(\mathbf{x}, t)),
    \end{equation}
    \begin{equation}\label{eqn:ddpmfwdkernel}
        q(\mathbf{\Psi}^{\prime}_{\tau}(\mathbf{x}, t)| \mathbf{\Psi}^{\prime}_{\tau-1}(\mathbf{x}, t)) = \mathcal{N}(\sqrt{1-\beta_{\tau}}\  \mathbf{\Psi}^{\prime}_{\tau-1}(\mathbf{x}, t), \beta_{\tau} \bm I).
    \end{equation}
\end{subequations}
The forward process given by the Markov chain and the transition kernel in equations~(\ref{eqn:ddpmfwd}) and (\ref{eqn:ddpmfwdkernel}) is known as the variance-preserving formulation. Note that $\beta_1, \beta_2, .. \beta_T$ denotes a monotonically increasing variance schedule with $0 < \beta_{\tau} < 1, \forall \tau \in \{1,\dots,T\}$. Furthermore, it follows that $q(\mathbf{\Psi}^{\prime}_{\tau}(\mathbf{x}, t)| \mathbf{\Psi}^{\prime}_{0}(\mathbf{x}, t)) = \mathcal{N}(\sqrt{\alpha_{\tau}}\ \mathbf{\Psi}^{\prime}_{0}(\mathbf{x}, t), (1-\alpha_{\tau}) \bm I)$ where $\alpha_{\tau} = \prod_{j=1}^{\tau} (1-\beta_j)$, allowing to directly sample $\mathbf{\Psi}^{\prime}_{\tau}(\mathbf{x}, t) \sim q( \mathbf{\Psi}^{\prime}_{\tau}(\mathbf{x}, t)| \mathbf{\Psi}^{\prime}_{0}(\mathbf{x}, t))$ for any $\tau \in [1, T]$ without iterating through the chain.
Similarly, the reverse process is a Markov chain with the reverse transition kernels approximated as Gaussian, see equations~(\ref{eqn:ddpmrev1}) and (\ref{eqn:ddpmrev2}).
\begin{subequations}
    \begin{equation}\label{eqn:ddpmrev1}
        p_\theta(\mathbf{\Psi}^{\prime}_{0:T}(\mathbf{x}, t)) = p(\mathbf{\Psi}^{\prime}_{T}(\mathbf{x}, t)) \textstyle\prod_{\tau=1}^T p_\theta(\mathbf{\Psi}^{\prime}_{\tau-1}(\mathbf{x}, t)| \mathbf{\Psi}^{\prime}_{\tau}(\mathbf{x}, t)),
    \end{equation}
    \begin{equation}\label{eqn:ddpmrev2}
        p_{\theta}(\mathbf{\Psi}^{\prime}_{\tau-1}(\mathbf{x}, t)|\mathbf{\Psi}^{\prime}_{\tau}(\mathbf{x}, t)) = \mathcal{N}(\bm \mu_{\theta}(\mathbf{\Psi}^{\prime}_{\tau}(\mathbf{x}, t), \tau), \bm\Sigma_{\phi}(\mathbf{\Psi}^{\prime}_{\tau}(\mathbf{x}, t),\tau)), \quad p(\mathbf{\Psi}^{\prime}_{T}(\mathbf{x}, t)) = \mathcal{N}(\bm 0, \bm I).
    \end{equation}
\end{subequations}
In equations~\ref{eqn:ddpmfwd} and~\ref{eqn:ddpmfwdkernel}, $q$ denotes the true forward transition kernel and in equations.~\ref{eqn:ddpmrev1} and~\ref{eqn:ddpmrev2}, $p_\theta$ denotes a parameterized reverse transition kernel with mean $\mu_{\theta}(\mathbf{\Psi}^{\prime}_{\tau}(\mathbf{x},t),\tau)$ and covariance $\bm\Sigma_{\theta}(\mathbf{\Psi}^{\prime}_{\tau}(\mathbf{x}, t),\tau)$, both of which are learned by neural networks with weights $\theta$ and $\phi$ respectively. In practice, we fix $\Sigma_{\phi}(\mathbf{\Psi}^{\prime}_{\tau} (\mathbf{x},t),\tau)=\beta_{\tau}\bm I$ and the mean is further parameterized as $\mu_{\theta}(\mathbf{\Psi}^{\prime}_{\tau}(\mathbf{x},t),\tau) = (1/\sqrt{1-\beta_{\tau}}) \left(\mathbf{\Psi}^{\prime}_{\tau}(\mathbf{x},t) - (\beta_{\tau}/ \sqrt{1-\alpha_{\tau}})\ \epsilon_{\theta}(\mathbf{\Psi}^{\prime}_{\tau}(\mathbf{x}, t),\tau) \right) $, where $\epsilon_{\theta}$ denotes the neural network. The diffusion model is trained using the variational objective which in turn simplifies to the $L_2$ loss between a randomly sampled standard normal noise $\epsilon_{\tau}$ and the model's prediction of this standard normal noise $\hat{\epsilon_{\tau}} = \epsilon_{\theta}(\mathbf{\Psi}^{\prime}_{\tau}(\mathbf{x}, t),\tau)$ from a noisy sample $\mathbf{\Psi}^{\prime}_{\tau}(\mathbf{x}, t) = \sqrt{\alpha_{\tau}} \mathbf{\Psi}^{\prime}_{0}(\mathbf{x}, t) + \sqrt{1-\alpha_{\tau}}\epsilon_{\tau}$, where $\mathbf{\Psi}^{\prime}_{0}(\mathbf{x}, t)$ denotes a clean data sample. For a detailed explanation of the DDPM training procedure, we refer to Refs.~\citep{ho2020, songDenoisingDiffusionImplicit2021}. During inference, starting from $\mathbf{\Psi}^{\prime}_{T}(\mathbf{x},t) \sim \mathcal{N}(\bm 0, \bm I)$, we can iteratively sample $\mathbf{\Psi}^{\prime}_{\tau-1}(\mathbf{x},t)$ from $\mathbf{\Psi}^{\prime}_{\tau}(\mathbf{x},t)$ until $\mathbf{\Psi}^{\prime}_{0}(\mathbf{x},t)$, which resembles a sample from the training data distribution.\\

The score-based generative modeling framework~\citep{song2021scorebased} connects the DDPM's learned mean prediction $\mu_{\theta}$ in terms of the score function of the intermediate distributions, i.e., $\nabla_{\mathbf{\Psi}^{\prime}_\tau} \log p(\mathbf{\Psi}^{\prime}_\tau)$, also known as the Stein score function. Consequently, the mean prediction $\mu_{\theta}$ can be equivalently written as:
\begin{equation}\label{eqn:scoreequivalence}
\mu_{\theta}(\mathbf{\Psi}^{\prime}_{\tau}(\mathbf{x},t),\tau) = \frac{1}{\sqrt{1-\beta_{\tau}}} \left(\mathbf{\Psi}^{\prime}_{\tau}(\mathbf{x},t) + \beta_{\tau} {s}_{\theta}(\mathbf{\Psi}^{\prime}_{\tau}(\mathbf{x}, t),\tau) \right),    
\end{equation} where ${s}_{\theta}(\mathbf{\Psi}^{\prime}_{\tau}(\mathbf{x}, t),\tau)$ is the neural network learned to approximate the true score function $\nabla_{\mathbf{\Psi}^{\prime}_{\tau} } \log p(\mathbf{\Psi}^{\prime}_\tau)$. 

In our experiments, we adopt the model architecture, the variance schedule, and other training parameters from the guided-diffusion repository~\footnote{\url{https://github.com/openai/guided-diffusion}}. We treat each snapshot as an independent and identically distributed (IID) sample of the underlying distribution of snapshots represented by the training set, a deliberate design choice suited to statistically stationary turbulence that enables conditioning on any sparse instantaneous observation without requiring temporally synchronized sensor measurements (see Supplementary Note~6 for detailed justification and validation).

\subsection*{Diffusion models for sparse reconstruction} \label{sec:method-reconstruction}

We turn our attention to the reconstruction problem described in equation~(\ref{eqn:invprob}), in which the goal is to reconstruct an instantaneous flow field $\mathbf{\Psi}^{\prime}(\mathbf{x}, t)$ from partial measurements $\mathcal{S}$ obtained through the forward operator $\mathcal{A}$ and potentially some measurement noise $\bm \eta \sim \mathcal{N}(\bm 0, \bm \sigma^2_s \bm I)$ with known variance $\sigma^2_s$. This general formulation of an inverse problem is a fundamental task with a host of different applications. In this work, we focus on the challenging case of noiseless inpainting, or sparse reconstruction, with the forward operator $\mathcal{A}(\mathbf{\Psi}^{\prime}(\mathbf{x},t)) = H \odot \mathbf{\Psi}^{\prime}(\mathbf{x},t)$, where $H$ denotes a binary measurement mask, with the measurement noise being zero, i.e. $\sigma_s = 0$.
\begin{equation}\label{eqn:invprob}
    \mathcal{S} = \mathcal{A}(\mathbf{\Psi}^{\prime}(\mathbf{x},t)) + \bm \eta .
\end{equation}
We solve the sparse reconstruction problem in a Bayesian setting using a pre-trained diffusion model as the prior and generate $\mathbf{\Psi}^{\prime}(\mathbf{x},t)$ (instantaneous flow) given the measurement $\mathcal{S}$ by sampling from the posterior distribution $p(\mathbf{\Psi}^{\prime}(\mathbf{x},t)| \mathcal{S} )$. This approach requires no additional training of the model and in fact, the same diffusion model can be used as a prior for all inverse problems given by equation~(\ref{eqn:invprob})~\citep{du_confild_2024, Li_S3GM_2024, Vishwasrao2024}. To this end, we explore two different methods: (1) $\Pi$GDM~\citep{songPseudoinverseGuidedDiffusionModels2022} and (2) MAPGA~\citep{Gutha2024InversePW}.\\

In equation~(\ref{eqn:scoreequivalence}), ${s}_{\theta}(\mathbf{\Psi}^{\prime}_{\tau}(\mathbf{x},t),\tau)$ learns the unconditional score function $\nabla_{\mathbf{\Psi}^{\prime}_{\tau}}\log p(\mathbf{\Psi}^{\prime}_{\tau})$ and sampling using this $\mu_{\theta}$ as in equation~(\ref{eqn:ddpmrev2}) results in sampling from $p(\mathbf{\Psi}^{\prime}_{0})$ as previously discussed. To solve an inverse problem, one needs to sample from the posterior distribution instead, i.e., from $p(\mathbf{\Psi}^{\prime}_{0}|\mathcal{S})$, where $\mathcal{S}$ denotes the measurements, see equation~(\ref{eqn:invprob}). For posterior sampling, it is equivalent to replacing the unconditional score function $\nabla_{\mathbf{\Psi}^{\prime}_{\tau}}\log p(\mathbf{\Psi}^{\prime}_{\tau})$ with the conditional score function $\nabla_{\mathbf{\Psi}^{\prime}_{\tau}}\log p(\mathbf{\Psi}^{\prime}_{\tau}|\mathcal{S})$ in equation~(\ref{eqn:scoreequivalence}), i.e, replacing ${s}_{\theta}(\mathbf{\Psi}^{\prime}_{\tau}(\mathbf{x},t),\tau)$ with another neural network ${s}_{\phi}(\mathcal{S}, \mathbf{\Psi}^{\prime}_{\tau}(\mathbf{x},t),\tau)$ approximating the conditional score function that is conditioned on the measurements $\mathcal{S}$. While learning a neural network that approximates this conditional score is possible, it requires additional training. Also, it remains task-specific, i.e., the model is specific to the given measurement operator and cannot be used for a different measurement operator.\ This remains a challenge to the SHAP framework, which inherently requires posterior sampling with different measurement operators to find the optimal sensor locations. In contrast, zero-shot methods such as $\Pi$GDM and MAPGA alleviate the need for additional training by only leveraging the unconditional model ${s}_{\theta}(\mathbf{\Psi}^{\prime}_{\tau}(\mathbf{x},t),\tau)$ for posterior sampling with arbitrary measurement operators, at the cost of performance. These zero-shot methodologies also fit well within the SHAP framework and are the focus of our sparse reconstruction pipeline in this work.
\begin{equation}\label{eqn:scoredecomposition}
    \nabla_{\mathbf{\Psi}^{\prime}_{\tau}}\log p(\mathbf{\Psi}^{\prime}_{\tau}|\mathcal{S}) = \nabla_{\mathbf{\Psi}^{\prime}_{\tau}}\log p(\mathbf{\Psi}^{\prime}_{\tau}) + \nabla_{\mathbf{\Psi}^{\prime}_{\tau}}\log p(\mathcal{S}|\mathbf{\Psi}^{\prime}_{\tau}), 
 \end{equation}
\begin{equation} \label{eqn:intractableterm}
    p(\mathcal{S}|\mathbf{\Psi}^{\prime}_{\tau}) = \int_{\mathbf{\Psi}^{\prime}_{0}} p(\mathcal{S}|\mathbf{\Psi}^{\prime}_{0})\ p(\mathbf{\Psi}^{\prime}_{0}|\mathbf{\Psi}^{\prime}_{\tau})\ \text{d}\mathbf{\Psi}^{\prime}_{0},
\end{equation}
\begin{equation} \label{eqn:tweediesformula}
    \mathbb{E}(\mathbf{\Psi}^{\prime}_{0}|\mathbf{\Psi}^{\prime}_{\tau}) = \frac{1}{\sqrt{\alpha_\tau}} \bigl( \mathbf{\Psi}^{\prime}_{\tau} + (1-\alpha_{\tau}) \nabla_{\mathbf{\Psi}^{\prime}_{\tau}}\log p(\mathbf{\Psi}^{\prime}_{\tau}) \bigr) \approx \frac{1}{\sqrt{\alpha_\tau}} \bigl( \mathbf{\Psi}^{\prime}_{\tau} + (1-\alpha_{\tau}){s}_{\theta}(\mathbf{\Psi}^{\prime}_{\tau},\tau) \bigr).
\end{equation}
In zero-shot methods, the conditional score can be decomposed as shown in equation~(\ref{eqn:scoredecomposition}), which consists of the unconditional score $\nabla_{\mathbf{\Psi}^{\prime}_{\tau}}\log p(\mathbf{\Psi}^{\prime}_{\tau})$ that can be substituted with ${s}_{\theta}(\mathbf{\Psi}^{\prime}_{\tau}(\mathbf{x},t),\tau)$ learned by the unconditional DDPM model. However, estimating $\nabla_{\mathbf{\Psi}^{\prime}_{\tau}}\log p(\mathcal{S}|\mathbf{\Psi}^{\prime}_{\tau})$ remains a challenge due to the intractable terms involved therein, see equation~(\ref{eqn:intractableterm}). $\Pi$GDM makes a Gaussian approximation of $p(\mathbf{\Psi}^{\prime}_{0}|\mathbf{\Psi}^{\prime}_{\tau}) = \mathcal{N}(\mu_{\tau},\sigma^2_{\tau}\bm{I})$, where $\mu_{\tau} = \mathbb{E}(\mathbf{\Psi}^{\prime}_{0}|\mathbf{\Psi}^{\prime}_{\tau})$, see equation~(\ref{eqn:tweediesformula}). The variance $\sigma^2_{\tau}$ is set proportional to $1-\alpha_{\tau}$. This approximation results in $p(\mathcal{S}|\mathbf{\Psi}^{\prime}_{\tau})$ being Gaussian and hence tractable. For additional details regarding the sampling, we refer to Ref.~\citep{songPseudoinverseGuidedDiffusionModels2022} on the $\Pi$GDM method.\\

MAPGA~\citep{Gutha2024InversePW} is yet another zero-shot diffusion posterior sampling method. However, in contrast to $\Pi$GDM, MAPGA circumvents approximating the conditional score function. Instead, it considers the deterministic mapping from the noisy sample $\mathbf{\Psi}^{\prime}_T$ to the clean sample $\mathbf{\Psi}^{\prime}_0$ given by the reverse Markov chain in DDPM (or equivalently, the Probability Flow path~\citep{song2021scorebased} in score-based generative modeling framework). With this mapping denoted as $f_{\theta}$, MAPGA solves the inverse problem by optimizing for the MAP estimate ($\mathbf{\Psi}^{\prime *}_0$) as shown in equations~(\ref{eqn:mapga_1}) and (\ref{eqn:mapga_2}). The optimization for $\mathbf{\Psi}^{\prime *}_T$ in equation~(\ref{eqn:mapga_1}) is performed via gradient ascent, as $\nabla_{\mathbf{\Psi}^{\prime}_T} \log p(\mathbf{\Psi}^{\prime}_0=f_{\theta}(\mathbf{\Psi}^{\prime}_T)|\mathcal{S})$ is tractable, see equations~(\ref{eqn:mapga_3}) and (\ref{eqn:mapga_4})~\citep{Gutha2024InversePW}. 
\begin{subequations}\label{eqn:mapga}
\begin{equation}\label{eqn:mapga_1}
\mathbf{\Psi}^{\prime *}_T = \arg\max_{\mathbf{\Psi}^{\prime}_T} \ \ \log p(\mathbf{\Psi}^{\prime}_0 = f_{\theta}(\mathbf{\Psi}^{\prime}_T)|\mathcal{S}),
\end{equation}
\begin{equation}\label{eqn:mapga_2}
\mathbf{\Psi}^{\prime *}_0 = f_{\theta}(\mathbf{\Psi}^{\prime *}_T),
\end{equation}
\begin{equation}\label{eqn:mapga_3}
\nabla_{\mathbf{\Psi}^{\prime}_T} \log p(\mathbf{\Psi}^{\prime}_0=f_{\theta}(\mathbf{\Psi}^{\prime}_T)|\mathcal{S}) = \frac{\partial f_{\theta}}{\partial \mathbf{\Psi}^{\prime}_T} \nabla_{\mathbf{\Psi}^{\prime}_0} \log p(\mathbf{\Psi}^{\prime}_0=f_{\theta}(\mathbf{\Psi}^{\prime}_T)|\mathcal{S}),
\end{equation}
\begin{equation}\label{eqn:mapga_4}
\nabla_{\mathbf{\Psi}^{\prime}_0} \log p(\mathbf{\Psi}^{\prime}_0=f_{\theta}(\mathbf{\Psi}^{\prime}_T)|\mathcal{S}) = \nabla_{\mathbf{\Psi}^{\prime}_0} \log p(\mathcal{S}|\mathbf{\Psi}^{\prime}_0=f_{\theta}(\mathbf{\Psi}^{\prime}_T)) + \nabla_{\mathbf{\Psi}^{\prime}_0} \log p(\mathbf{\Psi}^{\prime}_0=f_{\theta}(\mathbf{\Psi}^{\prime}_T)).
\end{equation}
\end{subequations}
Also, computing $\partial f_{\theta}/ \partial \mathbf{\Psi}^{\prime}_{\tau}$ involves backpropagating through the full reverse Markov chain, which is computationally infeasible. Instead, we approximate $f_{\theta}$ with the denoiser given by $\mathbf{E}(\mathbf{\Psi}^{\prime}_0|\mathbf{\Psi}^{\prime}_{\tau})$, see equation~(\ref{eqn:tweediesformula}), at each intermediate diffusion time step $\tau$. We follow the exact practical implementation detailed in MAPGA~\citep{Gutha2024InversePW}, with 20 diffusion time steps and 50 gradient ascent iterations per step. More implementation details can be found in the supplementary notes.

\subsection*{Mask design and segmentation}
\label{sec:mask-design}

Effective sparse reconstruction in real-world environments requires sensor placements that are not only informative but also practically deployable. Prior works based on linear methods, such as QR-pivoting~\citep{Manohar2017DataDrivenSS}, often identify high-energy grid points in the wake region; note that these zones are physically inaccessible or intrusive for actual sensor deployment. To address this, we design a modular binary mask that restricts candidate sensor locations to near-ground regions and the surfaces adjacent to the obstacle. Specifically, the mask includes all pixels satisfying:
\begin{equation}
  y/h \le d_{\mathrm{wall}} \quad \text{or} \quad \left| \mathbf{x} - \mathbf{x}_{\mathrm{wall}} \right| \le d_{\mathrm{wall}},
\end{equation}
with \( d_{\mathrm{wall}} = 0.3h \), thereby focusing placement on realistic areas close to structural surfaces and the ground.

The retained region comprises approximately 15\% of the total \(N_x \times N_y\) grid avoiding wake regions. For flexibility and modularity, the admissible region is subdivided into \(N_{\mathrm{seg}} = 900\) rectangular subregions, each of size \( 0.018\% \) of the total domain. These subregions serve as units of sensor selection in the SHAP attribution study. The size and form of a sensor can vary according to the practical use case, supporting scalable resolution and compatibility with real-world deployment constraints. These modular subregions serve as the atomic units for sensor attribution and selection in the SHAP-based optimization framework while decoupling placement logic from the raw DNS grid.
Thus, the segmentation flexibility enables a balance between spatial resolution and computational feasibility, particularly in the context of repeated reconstructions required by kernelSHAP.

\subsection*{Shapley values for optimal sensor placement}\label{sec:methods-shap}

SHAP (SHapley Additive exPlanations)~\citep{Lundberg} is a widely adopted framework for interpreting machine-learning models using Shapley values from cooperative game theory~\citep{shapley1953value}. In this study, we employ SHAP to evaluate the relative contribution of each candidate sensor subregion to the reconstruction accuracy achieved by our sparse reconstruction framework. 

Let the modular mask be defined as a set of \( M = 900 \) sensor-eligible subregions. We define a coalition \( \mathcal{C} \subseteq \{1, 2, \dots, M\} \) as a subset of active sensors. The reconstruction model, instantiated via MAPGA, generates a flow field \( \mathbf{\Psi}_0^{\prime\,(\mathcal{C})} \) conditioned on the sensor values \(\mathcal{S}_\mathcal{C}\). The value function:
\begin{equation}
    v(\mathcal{C}) = -\, \text{MSE}\left( \mathbf{\Psi}^{\prime}(\mathbf{x}, t),\; \mathbf{\Psi}_0^{\prime\,(\mathcal{C})} \right),
\end{equation}
where \( \mathbf{\Psi}^{\prime}(\mathbf{x}, t) \) is the ground-truth DNS field and \(\mathbf{\Psi}_0^{\prime\,(\mathcal{C})}\) is the MAP estimate generated by conditioning on coalition \(\mathcal{C}\). To estimate the marginal contribution of each sensor subregion \( i \), SHAP approximates \( v(\mathcal{C}) \) using a linear surrogate model over binary coalition indicators:
\begin{equation}
g(\mathcal{C}) \approx \phi_0 + \sum_{i=1}^{M} \phi_i \cdot \mathcal{C},
\label{eq:shap-additive}
\end{equation}
where \( \phi_i \) denotes the Shapley value assigned to subregion \(i\) and \(\phi_0\) represents the baseline value with no active sensors. These attributions are computed via the kernelSHAP algorithm~\citep{Lundberg}, which samples coalitions based on game theory and fits the additive model using weighted least squares:
\begin{equation}
\mathcal{L}(v, g, \pi) = \sum\left[ v(\mathcal{C}) - g(\mathcal{C}) \right]^2 \pi(\mathcal{C}),
\end{equation}
In the kernelSHAP framework, $\pi(\mathcal{C})$ denotes the weighting kernel used to guide the sampling of coalitions, effectively controlling the relative importance of different coalition sizes. Traditionally, SHAP has been applied to deterministic deep networks such as U-Nets~\citep{cremades_identifying_2024} and autoencoders, where the default kernel places higher weight on smaller coalitions to emphasize individual feature contributions. 
However, in probabilistic frameworks like ours, where reconstructions are guided by strong generative priors, the error--coverage relationship exhibits three distinct regimes, directly visible in figure~\ref{fig:shap-osp}(c)-(d). At very low coverage (below $k_{\min}$), all placement strategies converge to comparable, high reconstruction error: this is the prior-dominated regime, where the posterior is overwhelmingly governed by the diffusion prior rather than the sparse observations. Attributions computed in this regime systematically misattribute prior-driven reconstruction to individual sensors, producing importance maps that reflect the diffusion model's structure rather than genuine sensor utility. At intermediate coverage ($k_{\min}$ to $k_{\max}$), in the informative regime, sensor observations genuinely constrain the posterior beyond the prior. Beyond $k_{\max}$, the curves plateau into a saturation regime where marginal sensor utility approaches zero and additional attribution estimates are diluted by redundancy. The standard KernelSHAP kernel, which assigns disproportionately high weight to small coalitions, is therefore physically inappropriate in this setting. We therefore introduce a modified kernel $\pi_{\text{mod}}$ that restricts coalition sampling to this informative window, whose boundaries $k_{\min}$ and $k_{\max}$ are identified empirically, approximately from the inflection points visible in figure~\ref{fig:shap-osp}(c)-(d):
\begin{equation}
\pi_{\text{mod}}(\mathcal{C}) = 
\begin{cases}
\displaystyle \frac{M - 1}{\binom{M}{|\mathcal{C}|} \cdot |\mathcal{C}| (M - |\mathcal{C}|)} & \text{if } k_{\min} \leq |\mathcal{C}| \leq k_{\max}, \\
0 & \text{otherwise}.
\end{cases}
\end{equation}

Concretely, we set $k_{\min} = \lfloor 0.1 M \rfloor$ and $k_{\max} = \lfloor 0.4 M \rfloor$, corresponding to approximately 1.5--6\% pixel coverage of the total domain for the DNS dataset and 1--4\% pixel coverage for LT2400 dataset. The kernel is therefore not a freely tunable hyperparameter but a physically grounded design choice anchored in an observable feature of the reconstruction error--coverage relationship, and the truncated kernel ensures that coalitions outside this range are excluded. We compute Shapley values using a subset of snapshots sampled from the training dataset at a specific frequency for each dataset. For each snapshot, the MAPGA framework reconstructs the flow field using 3,000 sensor coalitions sampled according to the modified kernel \(\pi_{mod}\) based on cooperative game theory. Each coalition represents a distinct combination of active sensor subregions. The resulting Shapley values \(\phi_{i}\) quantify the marginal contribution of subregion \(i\) to reconstruction accuracy. The marginal contributions and coalition probabilities are then utilized to calculate the SHAP values. These SHAP values are then averaged across all sampled snapshots to construct a sensor importance map that reflects the expected utility of each subregion.

As discussed in the Results section, we evaluate the performance of SHAP value-based attribution by comparing it to two baseline methods: (1) a random baseline and (2) QR pivoting~\citep{Manohar2017DataDrivenSS}. In the random baseline, we generate seven random sensor configurations for each sensor count, ranging from 900 down to 20 subregions (corresponding to 15\% down to 0.3\% of the total domain area). The reconstruction MSE is evaluated for each mask and the resulting MSE vs.\% sensor coverage plot provides a performance envelope for unoptimized placements. For the QR pivoting method, we perform a pivoted QR decomposition on a subset of training snapshots, similar to SHAP attribution, yielding matrices \(Q\), \(R\) and \(P\). Here, \(Q\) represents an orthonormal basis capturing dominant modes, \(R\) is an upper-triangular matrix and \(P\) encodes column permutation based on importance. The ranking from \(P\) provides sensor prioritization and we retain only sensors that fall within the feasible baseline region. While QR pivoting captures linear structure, the corresponding reconstruction \(QRP\) yields poor results when applied directly to the masks discussed above, underscoring the importance of a learned diffusion prior for meaningful inference. For all strategies and sensor configurations, MAPGA is evaluated on the full test dataset using five different random seeds. The results are reported using two evaluation modes to assess the impact of different sources of uncertainty.

Together, these strategies allow us to assess the effectiveness and scalability of SHAP-based sensor placement under realistic deployment constraints.

\section*{Data availability}
The DNS dataset of the flow around a wall-mounted square cylinder at $\mathrm{Re}_h = 2000$ used in this study~\cite{MartinezSanchez2023} has been deposited in Zenodo under accession code \url{https://doi.org/10.5281/zenodo.22737509}.
The experimental PLIF concentration dataset of the LT2400 urban canopy~\cite{PLIF_ref} used in this study is available from the University of Southampton repository under accession code \url{https://doi.org/10.5258/SOTON/D2217}. Source data are provided with this paper.

\section*{Code availability}
The code for training the diffusion prior, MAPGA sparse reconstruction, SHAP-based sensor placement is publicly available at \url{https://github.com/VinuesaLAB-AI/Diff-SPORT}.

\section*{References}
\bibliography{Diff-SPORT-diffusion} 

\section*{Acknowledgements}
The authors thank Niki Loppi of the NVIDIA AI Technology Center Finland for valuable discussions and mentorship during the Poland Open Hackathon 2024, and Corentin Lapeyre, Strategic Researcher Engagement at NVIDIA, for his support and guidance.

\section*{Funding}
This work was supported by the University of Michigan; the Marie Sk\l{}odowska-Curie Actions project MODELAIR, funded by the European Union's HORIZON Research and Innovation Programme under grant agreement no.~101072559 (A.V., H.A., R.V.); the Swedish e-Science Research Centre (SeRC) and Digital Futures (H.A.); the Stanford Center for Turbulence Research under U.S. Office of Naval Research grant N000142312833 (R.V.); the European Union's HORIZON Research and Innovation Programme project RefMap, grant agreement no.~101096698 (R.V.); ERC grant no.~2021-CoG-101043998 DEEPCONTROL (R.V.); the Digital Futures programme at KTH (R.V.); and the NVIDIA Academic Grant Program (2025). Computations were enabled by resources provided by the National Academic Infrastructure for Supercomputing in Sweden (NAISS) on the Alvis system.

\section*{Author contributions}
A.V., S.B.C.G., K.W., H.A. and R.V. contributed to the ideation and design of the research; A.V. and S.B.C.G. developed the diffusion prior and the MAPGA reconstruction method; A.V., A.C. and R.V. developed the SHAP-based sensor placement framework; A.V. implemented the models and conducted the numerical experiments; A.V. and R.V. performed the DNS of the wall-mounted square cylinder and generated the high-fidelity data; H.D.L. and C.V. conducted the water-flume passive-scalar experiments and provided the experimental data; A.V. and R.V. analysed the data and results; A.V., S.B.C.G. and R.V. wrote the manuscript; A.P., C.G. and B.J.M. provided scientific input throughout the project and critically revised the manuscript; H.A. and R.V. supervised the project.

\section*{Competing interests}
The authors declare no competing interests.

\end{document}


\renewcommand{\thefigure}{S\arabic{figure}}
\renewcommand{\thetable}{S\arabic{table}}
\renewcommand{\theequation}{S\arabic{equation}}
\renewcommand{\thepage}{S\arabic{page}}
\begin{center}
    \LARGE\bfseries Supplementary Information for Diffusion-based sensor placement optimization and reconstruction of turbulent flows in urban environments
\end{center}
\vspace{1em}
\setcounter{page}{1}

\section*{Supplementary Note 1. MAPGA's effectiveness against DPS}

The paper compares MAPGA against the state-of-the-art zero-shot posterior sampling method $\Pi$GDM~\cite{songPseudoinverseGuidedDiffusionModels2022}. We would like to emphasize that Diffusion Posterior Sampling~\cite{chung2023diffusion} is identical to $\Pi$GDM in terms of formulation, except for a scaling factor. For completeness, we also evaluate DPS below on our sparse reconstruction task, and the results again demonstrate MAPGA's effectiveness over DPS.\\

\begin{figure}[h]
    \centering
    \includegraphics[width=0.7\textwidth]{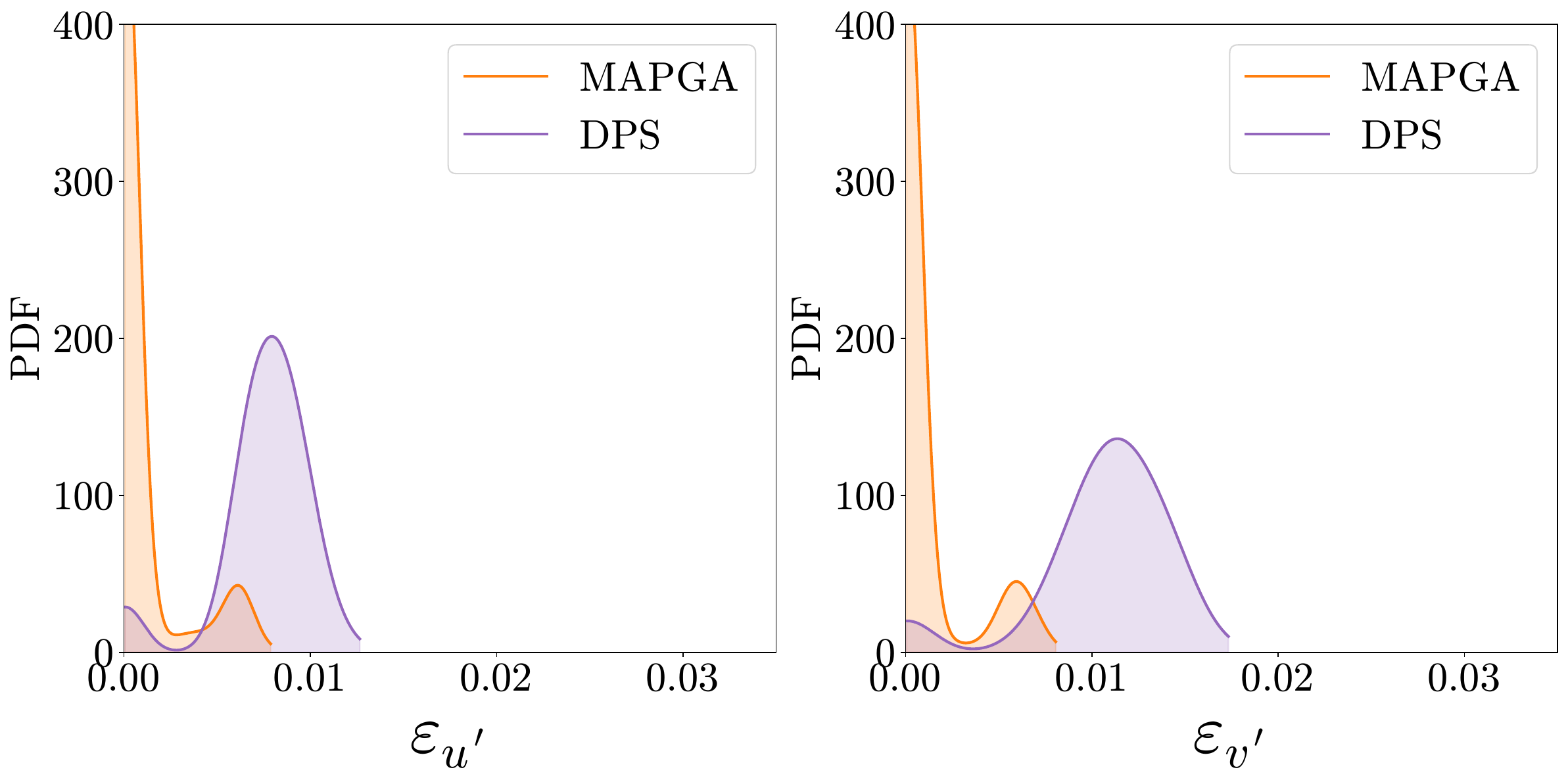}
    \caption{
    \textbf{Probability density functions (PDFs) of spatially averaged reconstruction error.}
    PDFs of \( \varepsilon \) for \( u^{\prime} \) (top) and \( v^{\prime} \) (bottom), comparing MAPGA and DPS. MAPGA consistently exhibits lower error compared to DPS, similar to its performance against \( \Pi\text{-GDM} \). Source data are provided as a Source Data file.
    }
    \label{fig:compare-dps}
\end{figure}

Note that posterior sampling methods like $\Pi$GDM and DDRM rely on estimating the conditional score $\nabla_{\mathbf{\Psi}^{\prime}_{\tau}}\log p(\mathbf{\Psi}^{\prime}_{\tau}|\mathcal{S})$ using Gaussian approximations and other heuristics. In contrast, MAPGA is a MAP-estimation method that aims to find the maximizer of the log-posterior (i.e., $\log p(\mathbf{\Psi}^{\prime}_0|\mathcal{S})$) that involves a different approach, and does not require estimating or approximating the conditional score. Therefore, it is naturally free from the Gaussian approximations or heuristics used by $\Pi$GDM, DPS, and DDRM. We believe MAPGA's efficacy stems from the fact that it explicitly aims to maximize the log-posterior, potentially generating highly likely samples of the posterior, unlike posterior sampling methods. Figure \ref{fig:compare-dps} shows the reconstruction error comparison between DPS and MAPGA for the test dataset.

\section*{Supplementary Note 2. Implementation details for training and inference}

\subsection{Diffusion model training}

\begin{figure}[h]
\begin{center}
    \includegraphics[width=0.8\textwidth,keepaspectratio]{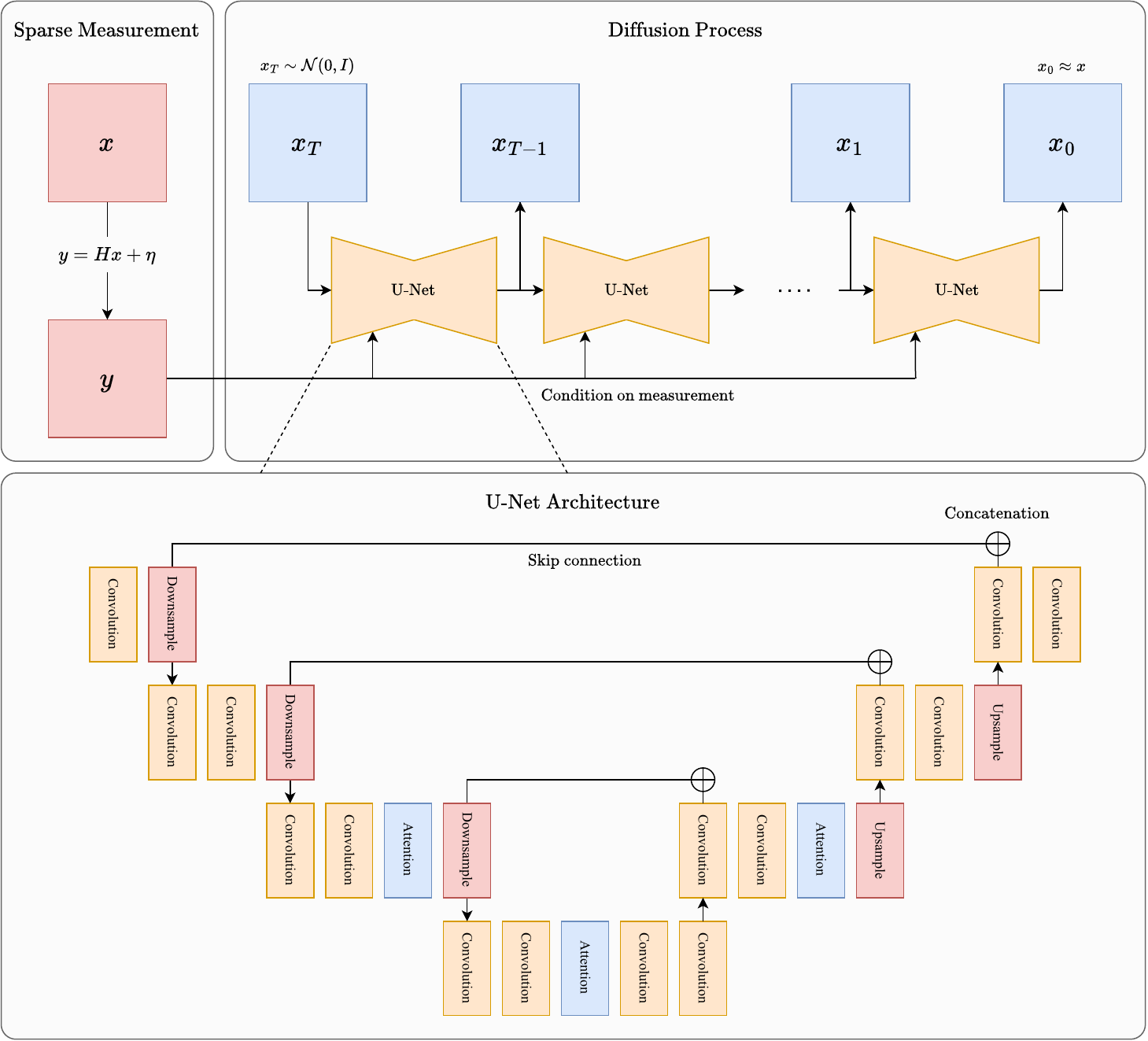}
    \caption{A general schematic of the sparse reconstruction process, combining the diffusion process with detailed layout of UNet architecture. At each diffusion timestep, the UNet receives a noisy image as input and outputs the denoised image.}
    \label{fig:schematic}
\end{center}
\end{figure}

For the diffusion model architecture, we adopt the same implementation as DDPM~\cite{ho2020}, with several important configuration choices adapted to our dataset, which differs significantly from standard image benchmarks. We employ a 2D U-Net architecture with an input image size of $(288, 96)$ and two input and output channels. The model uses $64$ base channels and consists of two residual blocks per resolution level. Attention is applied at spatial resolutions corresponding to downsampling factors of $4$ and $8$, with a single attention head. The channel multiplier is set to $(1, 2, 2, 2)$, and dropout is disabled. Convolutional resampling is enabled, gradient checkpointing is used to reduce memory consumption, and mixed-precision training (\texttt{fp16}) is not employed. The model predicts only the added noise and does not learn the reverse-process variances. Class conditioning and alternative attention ordering are disabled, and neither scale-shift normalization nor up/down residual blocks are used.

For diffusion training, we use a linear $\beta$ schedule with $\beta_{\text{start}} = 10^{-4}$ and $\beta_{\text{end}} = 0.02$, over $1000$ diffusion timesteps. The model is trained using the simplified loss $L_{\text{simple}}$ introduced in DDPM. Optimization is performed using the Adam optimizer with a learning rate of $5 \times 10^{-4}$ and a batch size of $32$. We do not employ an exponential moving average (EMA) of model parameters. Training is conducted for $1000$ epochs using our custom training code.

\subsection{MAPGA inference methodology}

We use the pre-trained model from above, and use the MAPGA~\cite{Gutha2024InversePW} inference method to reconstruct the full 2D flow snapshot $\mathbf{\Psi}^{\prime}_{0}(x,t)$ from given partial measurements $\mathcal{S} = H \odot \mathbf{\Psi}^{\prime}_{0}(x,t) + \eta $, where $H$ denotes the binary measurement matrix, and $\eta \sim \mathcal{N}(0, \sigma^2_s \mathbf{I})$.\\ 

Specifically, MAPGA aims to maximize the log-posterior, i.e., $\log p(\mathbf{\Psi}^{\prime}_0|\mathcal{S})$ at each diffusion time-step $\tau$, by considering the noise ($\mathbf{\Psi}^{\prime}_{\tau}$) to data ($\mathbf{\Psi}^{\prime}_0$) mapping approximated by the diffusion denoiser. Note that $\log p(\mathbf{\Psi}^{\prime}_0|\mathcal{S}) \propto \log p(\mathcal{S}|\mathbf{\Psi}^{\prime}_0) + \log p(\mathbf{\Psi}^{\prime}_0)$; which denotes the likelihood/data-consistency and the prior term, respectively.\\

Given a noisy sample $\mathbf{\Psi}^{\prime}_{\tau}$ at diffusion time $\tau$, generated by adding some standard normal noise $\hat{\epsilon}$ to some clean sample $\mathbf{\Psi}^{\prime}_{0}$ (i.e., $\mathbf{\Psi}^{\prime}_{\tau} = \sqrt{\alpha_{\tau}} \mathbf{\Psi}^{\prime}_{0} + \sqrt{1 - \alpha_{\tau}} \hat{\epsilon}$), the DDPM model $\epsilon_{\theta}(\mathbf{\Psi}^{\prime}_{\tau}, \tau)$ trained above learns to predict the added standard normal noise $\hat{\epsilon}$. By slight reparameterization of the model $\epsilon_{\theta}(\cdot, \cdot)$, it is equivalent to predicting the clean data sample $\mathbf{\Psi}^{\prime}_{0}$ instead, and this reparameterization is denoted by the denoiser $D_{\theta}(\mathbf{\Psi}^{\prime}_{\tau}, \tau)$. It can be shown that $\mathrm{D}_{\theta}(\mathbf{\Psi}^{\prime}_{\tau}, \tau) = \frac{1}{\sqrt{\alpha_t}} \bigl( \mathbf{\Psi}^{\prime}_{\tau} - \sqrt{1-\alpha_{\tau}} \epsilon_{\theta}(\mathbf{\Psi}^{\prime}_{\tau},\tau) \bigr)$. Considering this denoiser parameterization, we provide the MAPGA algorithm pseudo-code below.\\

\begin{figure}[h]
\begin{algorithm}[H]
\caption{\text{MAPGA}}
\label{algo:mapga}
$\ $\\
\KwInput{$\mathrm{D}_{\theta}(\cdot,\cdot), H, \mathcal{S}, \sigma_{s}, \alpha(\cdot), \sigma(\cdot), K, \lambda, \text{time-step schedule: }\{ \tau_N, \tau_{N-1}, \dots \tau_1, \tau_0 \},\text{ Note: } \sigma_{\tau_0} \ll 1$}
\KwOutput{$\mathbf{\Psi}^{\prime}_{\tau_0}$}

\KwInit{$\mathbf{\Psi}^{\prime}_{\tau_N} \sim \mathcal{N}(\mathbf{0},\mathbf{I})$}\\
$\ $\\

\For{$i \gets N$ \KwTo $1$}{
    $z \gets \mathbf{\Psi}^{\prime}_{\tau_i}$\\
    \For{$j \gets 1$ \KwTo $K$}{
        $\ $\\
        $\hat{\mathbf{\Psi}}^{\prime}_{\tau_0} = \mathrm{D}_{\theta}(z, \tau_i)$\DontPrintSemicolon\tcc*[r]{Noise to data mapping approximated by the denoiser}
        $\ $\\
        $\text{gradient}_{\text{likelihood}} = \frac{H^{\top}(\mathcal{S}-H \hat{\mathbf{\Psi}}^{\prime}_{\tau_0})}{\sigma^2_s + \sigma^2_{\tau_0}}$\DontPrintSemicolon\tcc*[r]{Gradient of the data consistency term}
        $\ $\\
        $\text{gradient}_{\text{prior}} = \frac{ \mathrm{D}_{\theta}(\hat{\mathbf{\Psi}}^{\prime}_{\tau_0}, \tau_0) - \hat{\mathbf{\Psi}}^{\prime}_{\tau_0}}{\sigma^2_{\tau_0}}$\DontPrintSemicolon\tcc*[r]{Gradient of the prior term}
        $\ $\\
        $\text{gradient}_{\text{posterior}} = \frac{\partial \mathrm{D}^{\top}_{\theta}(z,\tau_i)}{\partial z} (\text{gradient}_{\text{likelihood}} + \text{gradient}_{\text{prior}})$\DontPrintSemicolon\tcc*[r]{Gradient of the posterior w.r.t z \\ See Equation (10c) in the paper}
        $\ $\\
        $z \gets z + \lambda \cdot \text{gradient}_{\text{posterior}}$\DontPrintSemicolon\tcc*[r]{Gradient ascent update for optimal z}
        $\ $\\
        
    }
    $\mathbf{\Psi}^{\prime}_{\tau_{i-1}} \sim \mathcal{N}(\sqrt{\alpha}_{\tau_{i-1}} \mathrm{D_{\theta}}(z,\tau_i), (1 - \alpha_{\tau_{i-1}})\mathbf{I})$\DontPrintSemicolon\tcc*[r]{Map optimal z to data and stochastic resample}
}
\KwRet{$\mathbf{\Psi}^{\prime}_{\tau_0}$}

\end{algorithm}
\end{figure}

The inputs to the algorithm above include the denoiser $\mathrm{D}_{\theta}(\cdot, \cdot)$, the binary measurement matrix $H$, the measurements $\mathcal{S}$, the measurement noise $\sigma_s$, the number of reverse diffusion steps $N$, the corresponding time-step schedule $\{\tau_N, \tau_{N-1}, \dots \tau_1, \tau_0\}$, the number of gradient-ascent iterations per step $K$, and the learning rate $\lambda$. Note that $\alpha(\cdot)$ denotes the variance schedule of the pre-trained DDPM model, and we define $\sigma(\cdot) = \sqrt{\frac{1-\alpha(\cdot)}{\alpha(\cdot)}}$. For convenience, we use the notation $\alpha_{\tau}$ and $\sigma_{\tau}$ instead to denote $\alpha(\tau)$ and $\sigma(\tau)$ in the algorithm above.\\

In practice, we use the default hyperparameters of MAPGA, i.e., $N=20$, $K=50$, $\lambda = \sigma^2_s + \sigma^2_{\tau_0}$. Though it may improve performance, we do not set any explicit weights to the likelihood and prior terms since it introduces additional hyperparameters. Also, in the original paper~\cite{Gutha2024InversePW}, MAPGA by default used a time-step schedule corresponding to the EDM~\cite{karras2022elucidating} noise schedule. However, we instead used a uniform time-step schedule in all our experiments in the paper, with $\tau_N = 1000$ and $\tau_0 = 1$. Later, we found that setting the MAPGA's time-step schedule back to its default, which corresponds to the EDM noise schedule, i.e., $\tau_i = \sigma^{-1}(\sigma_{\tau_i})$, where $\sigma_{\tau_i} = {({\sigma_{min}}^{\frac{1}{\rho}} + \frac{i}{N} ({\sigma_{max}}^{\frac{1}{\rho}} - {\sigma_{min}}^{\frac{1}{\rho}}) )}^{\rho}$, with $\sigma_{min}=0.002,\ \sigma_{max}=140$ and $\rho=7$, performs even better, as described in the later sections. The gradient ascent runs for a fixed $K$ iterations per diffusion step with no early stopping criterion. The reverse diffusion transitions (outer loop, Algorithm~1, line~9) use stochastic DDPM sampling, Gaussian noise is injected at each timestep transition, rather than deterministic DDIM steps. The final model checkpoint is used directly at inference without exponential moving average (EMA). No explicit regularization is applied to the MAP objective beyond the implicit regularization provided by the diffusion prior.\\

\section*{Supplementary Note 3. Robustness and generalization checks}

\subsection{Reconstruction under measurement noise}

To assess MAPGA's robustness to noisy observations, we evaluate reconstruction performance under varying signal-to-noise ratios (SNR). The SNR quantifies measurement quality as the ratio of signal power to noise power within the observed region $\Omega$. Specifically, we define SNR in decibels as
\begin{equation}
\text{SNR}_{\text{dB}} = 10\log_{10}\!\left(\frac{\frac{1}{|\Omega|}\sum_{i\in\Omega} \mathcal{S}_i^2}{\sigma^2}\right),
\end{equation}
where $\mathcal{S}_i$ denotes the observed signal at location $i$ and $\sigma^2$ is the noise variance. Lower SNR values correspond to noisier measurements: at 10~dB, noise power is 10\% of signal power, representing challenging but realistic sensor conditions, while 25~dB reflects high-quality measurements with minimal noise contamination.\\

We test MAPGA across three OSP strategies, SHAP-based, QR-pivoting, and random sensor placement—under SNR levels ranging from 10 to 25~dB. For each configuration, we add Gaussian noise $\eta \sim \mathcal{N}(0, \sigma^2 \mathbf{I})$ to the observed measurements $\mathcal{S} = H \odot \mathbf{\Psi}^{\prime}_0 + \eta$ and reconstruct the full flow field using MAPGA with the nominal (noise-free) measurement operator $H$.\\

\begin{figure}[h]
    \centering
    \includegraphics[width=0.7\textwidth]{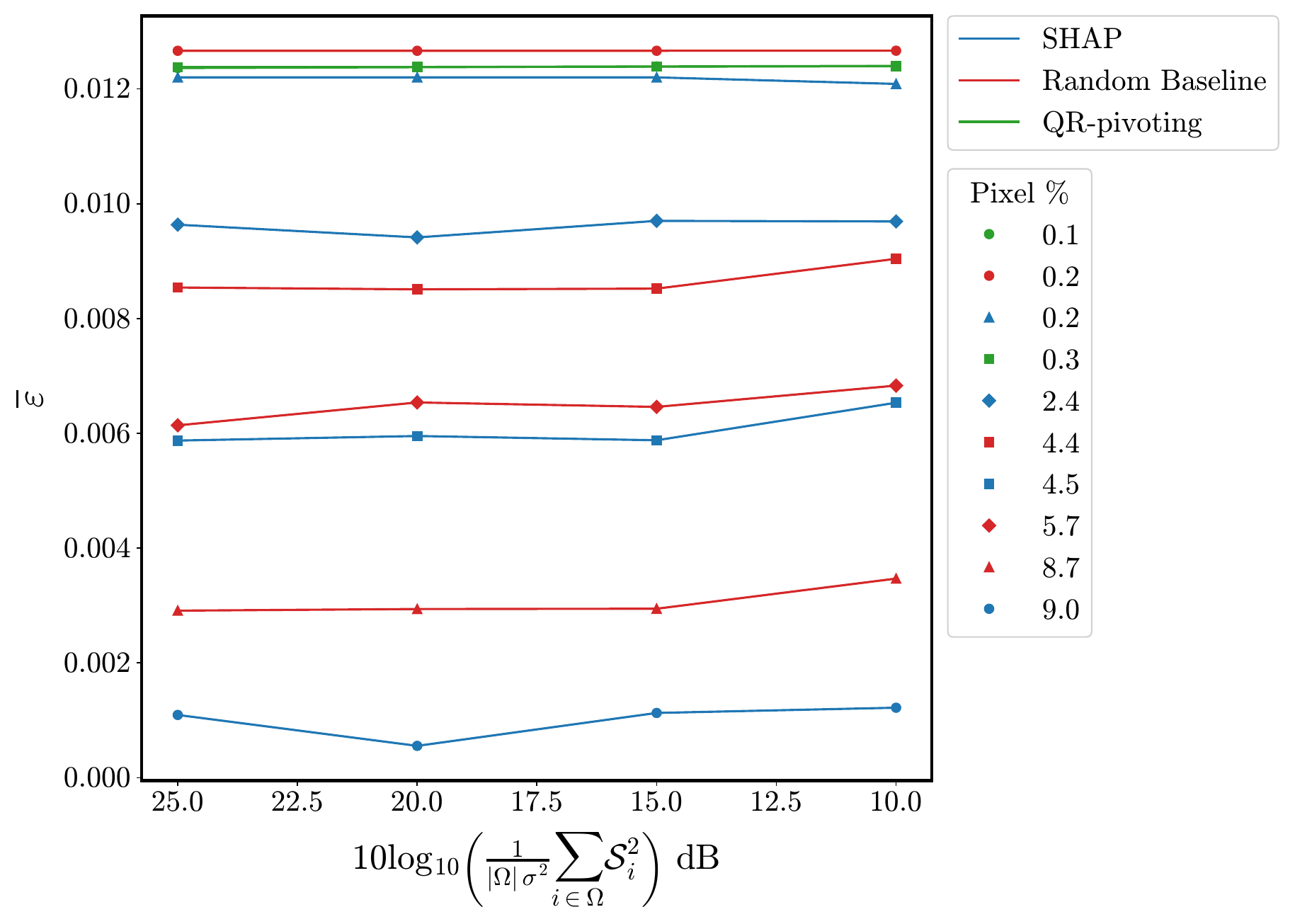}
    \caption{
    \textbf{Reconstruction error versus signal-to-noise ratio (SNR).}
    Average reconstruction error $\overline{\varepsilon}$ ($n = 20$) plotted against SNR for three OSP methods: SHAP-based, random baseline, and QR-pivoting. Different markers indicate varying sensor coverage levels (pixel \%). SHAP-based placement exhibits superior noise robustness, maintaining lower error across all SNR levels. The x-axis is inverted so that noise severity increases from left to right. Source data are provided as a Source Data file.
    }
    \label{fig:snr-robustness}
\end{figure}

Figure~\ref{fig:snr-robustness} shows the reconstruction error $\overline{\varepsilon}$ as a function of SNR for different OSP methods and sensor coverage levels (indicated by pixel percentage). For SNR values above 15~dB, MAPGA exhibits minimal degradation in reconstruction quality across all OSP methods, with error remaining nearly constant and demonstrating effective handling of moderate measurement noise. As SNR decreases below 15~dB (increasing noise levels), reconstruction error increases modestly; importantly, this degradation is gradual rather than abrupt, indicating that MAPGA does not suffer from catastrophic failure under noisy conditions. The choice of sensor placement strategy has a pronounced impact on noise robustness: random sensor placement (red curves) shows steeper error growth as SNR decreases compared to SHAP-based placement (blue curves), with SHAP achieving 50--70\% lower reconstruction error than random placement at comparable sensor coverage (4--9\% of pixels) across all noise levels. SHAP-based OSP consistently maintains the lowest reconstruction error across the entire SNR range, and notably, even at high sensor coverage (9.0\%), SHAP outperforms QR-pivoting and random strategies, demonstrating that optimized sensor placement provides inherent robustness to measurement noise.

These results demonstrate that MAPGA is robust to realistic measurement noise, with reconstruction quality degrading gracefully as SNR decreases. Critically, the choice of OSP method—particularly SHAP-based placement plays a more significant role in determining reconstruction accuracy than the noise level itself within the tested range. This suggests that investing in optimized sensor placement yields greater practical benefits than pursuing marginally lower sensor noise in real-world deployments.

\subsection{Robustness to sensor misregistration}

In real-world deployments, sensor positions may deviate from their nominal locations due to installation errors, thermal expansion, vibration, or calibration drift. To evaluate MAPGA's robustness to such operator mismatch, we test a challenging scenario where observations are collected using perturbed sensor positions, but reconstruction is performed assuming nominal (ideal) positions. This setup deliberately creates a mismatch between the forward operator used to generate observations ($H_{\text{obs}}$) and the operator used in MAPGA's reconstruction ($H_{\text{rec}}$), representing the worst-case deployment scenario where misregistration goes undetected.\\

We apply spatial shifts (1--2 pixels in $x$ and/or $y$ directions) and rotations (0.5$^{\circ}$) to SHAP-based sensor placements at three coverage levels (corresponding to approximately 9\%, 4.5\%, and 2.4\% pixel coverage). Each perturbation type is tested independently (applied to all sensor locations), along with a combined perturbation scenario that applies both 1-pixel diagonal shift and 0.5$^{\circ}$ rotation simultaneously. Table~\ref{tab:misregistration} summarizes the reconstruction errors for these configurations.\\

\begin{table}[h]
\centering
\caption{Reconstruction error under sensor misregistration. SHAP masks at three coverage levels are tested under various perturbations. The aligned configuration denotes no perturbation (baseline). MSE values show that small misalignments (1 px, 0.5$^{\circ}$) cause modest degradation, while 2-pixel shifts significantly degrade performance, particularly for sparser masks.}
\label{tab:misregistration}
\begin{tabular}{lcccccc}
\toprule
& \multicolumn{2}{c}{\textbf{Perturbation}} & \multicolumn{3}{c}{\textbf{$\overline{\varepsilon}$ ($\times 10^{-3}$)}} \\
\cmidrule(lr){2-3} \cmidrule(lr){4-6}
\textbf{Configuration} & Shift (px) & Rotation ($^{\circ}$) & 9.0\% Pixels & 4.5\% Pixels & 2.4\% Pixels \\
\midrule
Aligned (baseline)     & 0       & 0.0   & 1.09  & 5.77  & 10.14 \\
1-pixel diagonal shift & 1       & 0.0   & 1.18  & 5.70  & 9.51  \\
0.5$^{\circ}$ rotation & 0       & 0.5   & 1.73  & 6.53  & 9.90  \\
2-pixel shift          & 2       & 0.0   & 7.76  & 10.27 & 11.69 \\
Combined (shift + rot) & 1       & 0.5   & 1.13  & 6.02  & 10.00 \\
\bottomrule
\end{tabular}
\end{table}

Small perturbations representative of worst realistic deployment conditions, 1-pixel shifts or 0.5$^{\circ}$ rotations to all sensor locations, induce only modest increases in reconstruction error. For the high-coverage SHAP mask (9\% pixels), a 1-pixel diagonal shift increases MSE by less than 10\% (from 1.09 to 1.18 $\times 10^{-3}$), while 0.5$^{\circ}$ rotation increases error by approximately 60\% (to 1.73 $\times 10^{-3}$), still maintaining acceptable reconstruction quality. This demonstrates that MAPGA tolerates typical sensor positioning errors encountered in practical experimental setups.

Larger misalignments, 2-pixel shifts, cause significant performance degradation, with MSE increasing by factors of 2$\times$ to 7$\times$. This sensitivity to larger perturbations is expected, as 2-pixel misregistration at the native grid resolution represents a substantial deviation that violates the assumption of accurate operator knowledge. Importantly, even under this severe mismatch, MAPGA does not fail catastrophically but degrades gracefully. The impact of misregistration depends on sensor coverage: sparser masks (lower thresholds) exhibit higher baseline errors and show different sensitivity patterns. The combined perturbation scenario (1-pixel shift + 0.5$^{\circ}$ rotation) yields errors comparable to individual 1-pixel shifts, suggesting that the effects of small simultaneous perturbations do not compound dramatically.

These results demonstrate that MAPGA is robust to realistic sensor misregistration errors typically encountered in experimental deployments. While maintaining strict calibration is desirable, the method's graceful degradation under moderate operator mismatch substantiates its practical utility in real-world flow reconstruction applications where perfect sensor positioning cannot be guaranteed.

\subsection{Performance under out-of-distribution variation}

To evaluate the robustness of Diff--SPORT under out-of-distribution (OOD) conditions, we assess reconstruction performance on flow data extracted from spanwise locations that were not used during training of the diffusion prior. The DDPM prior is trained exclusively on two-dimensional velocity snapshots extracted at the spanwise plane $z/h = 0$ from the three-dimensional DNS dataset. All main results in this work are obtained using this training distribution.

For the OOD test, the trained diffusion prior is kept fixed and the full downstream reconstruction pipeline, comprising the learned prior, the measurement operator, and MAPGA posterior inference, is applied to flow snapshots extracted at unseen spanwise planes $z/h = 1/8$ and $z/h = 1/4$. It must be noted that the instantaneous flow statistics differ across spanwise planes due to the highly turbulent three-dimensional nature of the flow. This setup therefore constitutes a realistic but significant distribution shift within the same flow configuration.

Table~\ref{tab:ood_spanwise} reports the mean-squared reconstruction error for the streamwise ($u$) and wall-normal ($v$) velocity components at each spanwise location. As expected, the reconstruction error increases moderately with increasing distance from the training plane.

Note that MAPGA continues to reconstruct physically plausible flow fields with bounded error across all the tested spanwise locations. Moreover, the reconstruction errors obtained via MAPGA remain substantially lower than those produced by unconditional samples drawn from the diffusion prior alone, indicating that measurement-conditioned posterior inference mitigates the effects of distribution shift. These results demonstrate robustness to modest OOD variation within a statistically stationary turbulent flow, while highlighting the corrective role of conditioning relative to the learned prior.

At the same time, these findings reinforce the central modeling assumption of this work: Diff--SPORT is designed for statistically stationary flows, and reconstruction performance degrades progressively as inference moves further away from the training distribution. Larger spanwise offsets, strongly non-stationary dynamics, or changes in flow configuration would likely require multi-plane training or explicitly conditional priors, which are beyond the scope of the present study.

\begin{table}[h]
\centering
\caption{Reconstruction error under modest out-of-distribution variation across spanwise planes. The diffusion prior is trained exclusively on data from $z/h=0$ and applied without retraining to unseen spanwise locations. Errors are reported separately for streamwise ($u$) and wall-normal ($v$) velocity components.}
\label{tab:ood_spanwise}
\begin{tabular}{lcc}
\toprule
\textbf{Spanwise location} & \textbf{$\varepsilon_{u^{\prime}}$} & \textbf{$\varepsilon_{v^{\prime}}$ }\\
\midrule
$z/h = 0$ (in-distribution) & 7.86$\times 10^{-4}$ & 8.70$\times 10^{-4}$ \\
$z/h = 1/8$                & 6.08$\times 10^{-3}$ & 7.73$\times 10^{-3}$ \\
$z/h = 1/4$                & 1.24$\times 10^{-2}$ & 1.14$\times 10^{-2}$ \\
\bottomrule
\end{tabular}
\end{table}

\begin{figure}[h]
    \centering
    \includegraphics[width=0.9\textwidth]{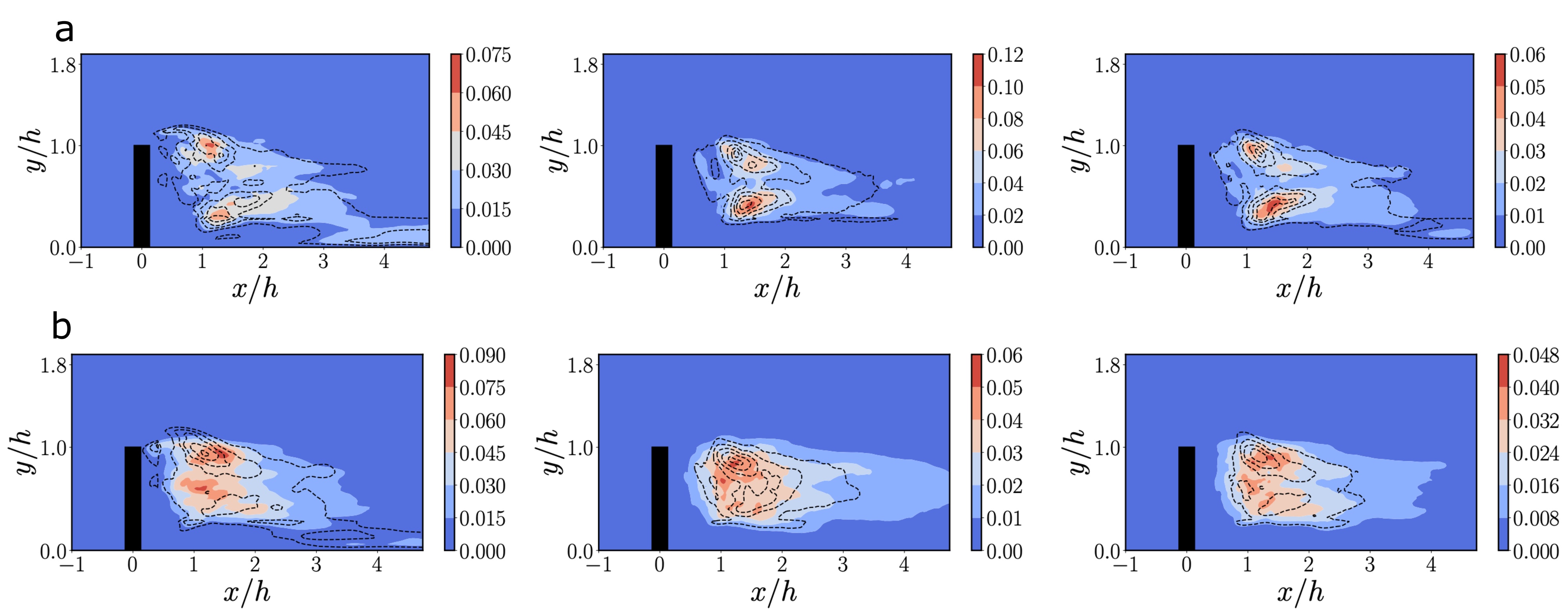}
    \caption{
    \textbf{Reynolds stresses on out-of-distribution spanwise planes.}
    Filled contours: DNS ground truth. Dashed contours: MAPGA reconstruction. Columns: $\overline{u^{\prime}u^{\prime}}$, $\overline{v^{\prime}v^{\prime}}$, $\overline{u^{\prime}v^{\prime}}$. (a) $z/h=1/8$; (b) $z/h=1/4$.
    }
    \label{fig:ood_reynolds}
\end{figure}

To characterize this degradation beyond the scalar MSE values reported above, Figure~\ref{fig:ood_reynolds} shows the second-order statistics, Reynolds stress fields ($\overline{u^{\prime}u^{\prime}}$, $\overline{v^{\prime}v^{\prime}}$, $\overline{u^{\prime}v^{\prime}}$), computed from the full reconstructed ensemble at each OOD plane, with the DNS ground truth shown as filled contours and MAPGA reconstructions overlaid as dashed contours. At $z/h=1/8$ (panel a), the reconstructed stress fields closely track the ground-truth spatial structure, including the bilobed shear-layer pattern straddling the wake centerline. At $z/h=1/4$ (panel b), the larger distribution shift from the training plane is reflected in a visibly coarser stress field with reduced peak magnitude relative to DNS, consistent with the higher MSE reported in Table~\ref{tab:ood_spanwise}. In both cases, MAPGA reproduces the dominant spatial organization of the turbulent stresses without diverging or producing unphysical artifacts, indicating that the degradation with distance from the training plane is graceful.

\section*{Supplementary Note 4. SHAP sensitivity analysis}

\subsection{Sensitivity to coalition size}

SHAP attribution requires evaluating a scoring function, for a random sample of coalition masks. The number of coalitions $n_{\text{coali}}$ controls the fidelity of the attribution estimates: too few coalitions produce noisy importance maps, while more coalitions increase computational cost proportionally. We evaluate sensitivity to this parameter by repeating the full SHAP computation at four coalition budgets: $n_{\text{coali}} \in \{430, 858, 2145, 3000\}$, corresponding to approximately $1\times$, $2\times$, $5\times$, and $7\times$ the number of valid pixels within the feasibility mask or valid features available for coalition formation (429 pixels).

\begin{figure}[h]
    \centering
    \includegraphics[width=0.65\textwidth]{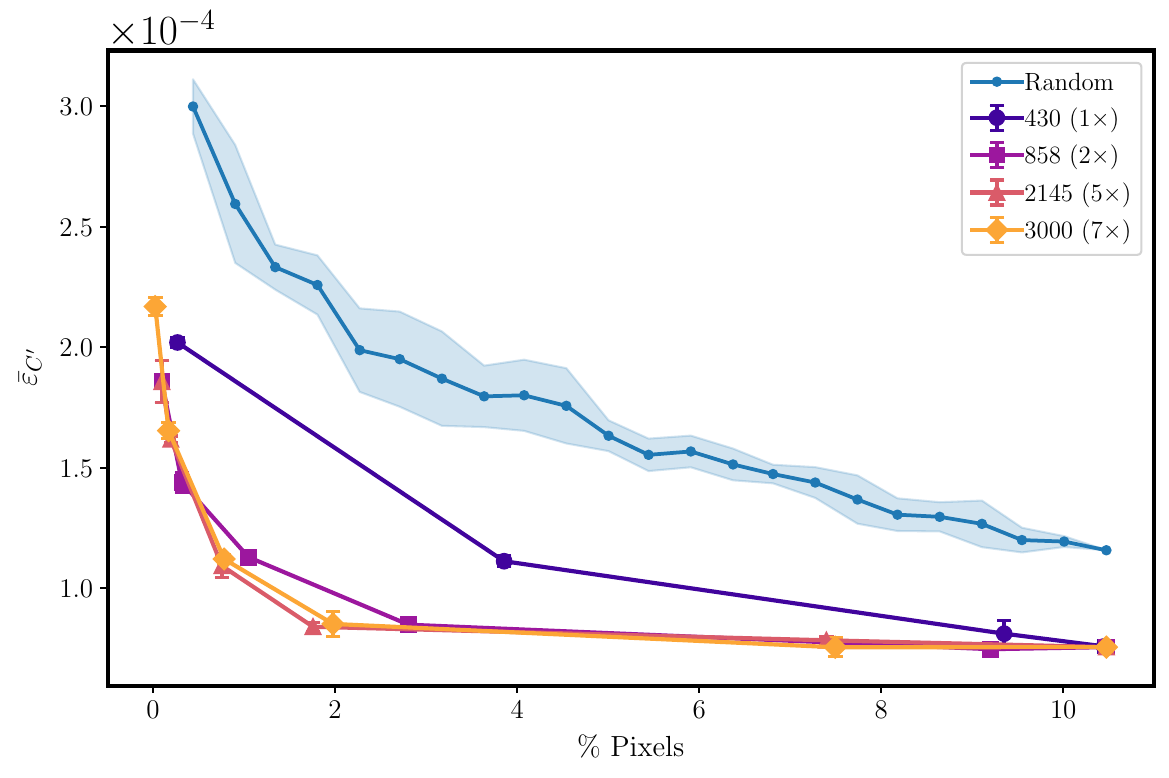}
    \caption{
    \textbf{SHAP attribution robustness to coalition size.}
    Average reconstruction error $\bar{\varepsilon}_{C'}$ vs.\ sensor coverage (\% pixels) for four coalition sizes ($n_{\text{coali}} \in \{430, 858, 2145, 3000\}$) and a random placement baseline. Error bars denote $\pm$ one standard deviation across 4 seeds for each coalition size. SHAP performance is stable across the full coalition range, confirming insensitivity to this parameter above ${\sim}1\times$ the feature count. Source data are provided as a Source Data file.
    }
    \label{fig:coalition_ablation}
\end{figure}

Figure~\ref{fig:coalition_ablation} shows the average reconstruction error $\bar{\varepsilon}_{C'}$ as a function of sensor coverage for all four coalition sizes, together with the random placement baseline. SHAP attribution is robust to coalition size: all four variants yield consistent error curves, and even the budget configuration ($n_{\text{coali}} = 430$) showcases lower error than the random placement errors. This confirms that SHAP offers robust sensor placement for all the available coalition counts as compared to random baseline, however coalition count remains a critical sensitivity parameter affecting the computational costs and marginal impact on accuracy.

\subsection{Sensitivity to segmentation granularity}

The SHAP feature space is defined by superpixel segmentation of the feasibility mask, where a block size of $b \times b$ pixels groups spatially adjacent pixels into a single feature. Pixel-level attribution ($b=1$) treats each valid pixel independently (429 features), while coarser segmentations ($b=2$, $b=4$) reduce the feature count to approximately 184 and 66 features, respectively. We evaluate whether this design choice affects OSP quality.

\begin{figure}[h]
    \centering
    \includegraphics[width=0.9\textwidth]{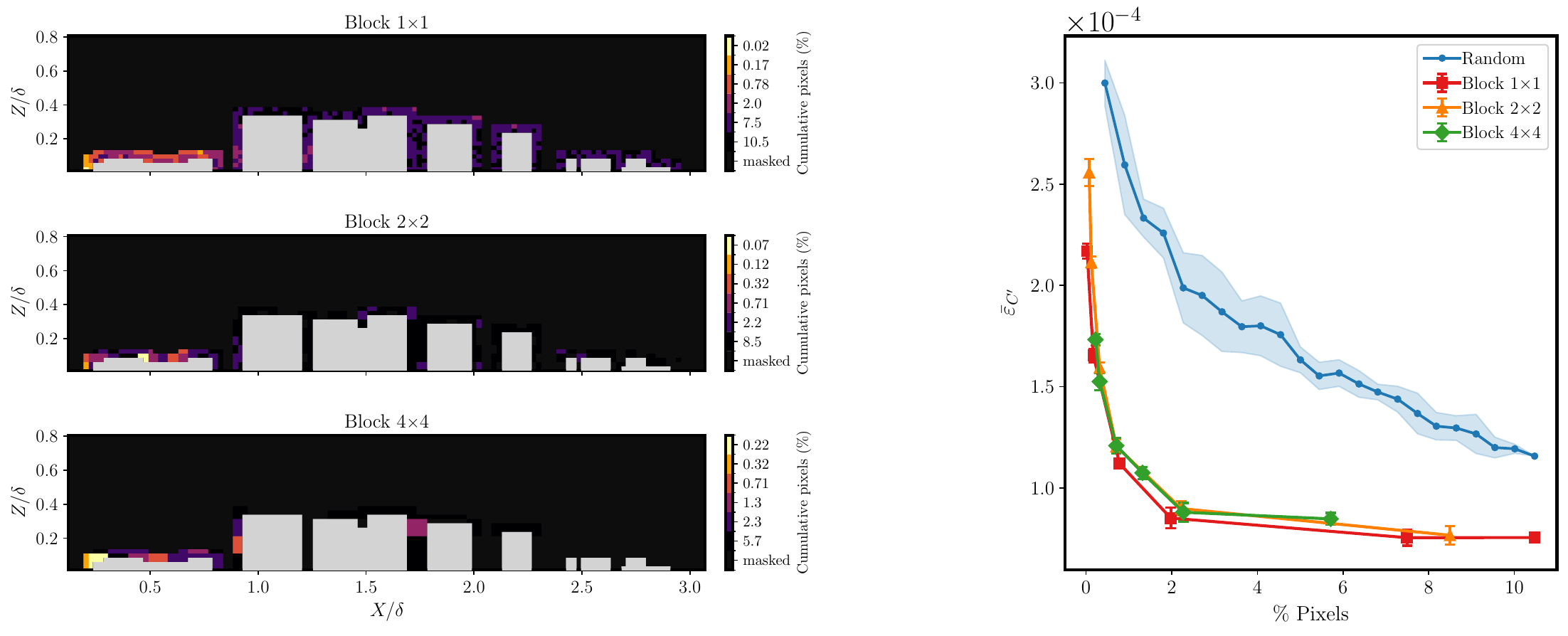}
    \caption{
    \textbf{SHAP attribution robustness to segmentation granularity.}
    Left: SHAP importance maps for block sizes $1\times1$ (pixel-level), $2\times2$, and $4\times4$, shown at six sensor coverage thresholds. Per-variant cumulative coverage percentages are annotated. Right: Reconstruction error $\bar{\varepsilon}_{C'}$ vs.\ sensor coverage for the three block sizes and random baseline. Error bars denote $\pm$ one standard deviation across 4 seeds ($n = 260$). Pixel-level attribution ($b=1$) achieves marginally lower error; coarser blocks trade a small accuracy loss for reduced feature dimensionality and proportionally lower coalition budgets, offering a practical accuracy--speed tradeoff. Source data are provided as a Source Data file.
    }
    \label{fig:segmentation_ablation}
\end{figure}

Figure~\ref{fig:segmentation_ablation} shows reconstruction error curves for block sizes $b \in \{1, 2, 4\}$. Pixel-level attribution ($b=1$) achieves marginally lower reconstruction error than coarser segmentations across the full sensor coverage range, as finer resolution better resolves spatially adjacent but informationally distinct sensor locations. The differences are small at the current grid resolution but are expected to grow at higher resolutions, where coarser superpixels increasingly average over heterogeneous flow structures.

The key practical implication is an accuracy, speed tradeoff mediated by feature count. Increasing block size from $1{\times}1$ to $4{\times}4$ reduces the number of SHAP features from 429 to approximately 66 (${\sim}6.5\times$ reduction). Since SHAP convergence requires a coalition budget that scales with feature count, at production quality ($7\times$ the feature count), block $4{\times}4$ requires only ${\sim}462$ coalitions versus ${\sim}3000$ for block $1{\times}1$, wall-clock time reduces proportionally (see Table~\ref{tab:wallclock}). Users may therefore adopt coarser segmentation to trade a small accuracy loss for substantially faster OSP design, particularly useful in high-resolution domains or compute-constrained deployments.

\section*{Supplementary Note 5. Computational cost of the Diff--SPORT pipeline}
\label{sec:wallclock}

Table~\ref{tab:wallclock} reports wall-clock times for all three stages of the Diff--SPORT pipeline across both experimental datasets: the DNS velocity dataset ($288{\times}96$ grid, 2 channels: $u'$, $v'$) and the PLIF concentration dataset ($128{\times}32$ grid, 1 channel: $C'$). All times are normalised per output channel to enable a direct comparison across datasets with different channel counts. Training times are an exception: since all channels are trained simultaneously in a single model, training time is reported as a single total and is not divided per channel.

\begin{table}[h]
\centering
\caption{
\textbf{Wall-clock times for the Diff--SPORT pipeline across both datasets, reported per channel.}
Training times are totals (all channels trained simultaneously in one model). MAPGA and SHAP-OSP times are per snapshot per channel. SHAP-OSP times are additionally per generation seed. SHAP ablation rows (coalition-size and block-size variants) were evaluated on the PLIF dataset only, the DNS entry reports the production configuration ($n_{\text{coali}}=3000$, block $1\times1$).
}
\label{tab:wallclock}
\begin{tabular}{llcc}
\toprule
\textbf{Stage} & \textbf{Configuration} & \textbf{Hardware} & \textbf{Wall-clock (per channel)} \\
\midrule
\multicolumn{4}{l}{\textit{Prior training (one-time)}} \\[2pt]
DDPM training (DNS, 2 ch.)  & 1000 epochs & 4$\times$ A100 & 44\,h\,47\,min (total) \\
DDPM training (PLIF, 1 ch.) & 1000 epochs & 1$\times$ A100 & 2\,h\,33\,min (total) \\[4pt]
\midrule
\multicolumn{4}{l}{\textit{MAPGA reconstruction (per snapshot)}} \\[2pt]
MAPGA inference (DNS)  & $N=20$, $K=50$ & 1$\times$ A100 & $\approx$6\,s \\
MAPGA inference (PLIF) & $N=20$, $K=50$ & 1$\times$ A100 & $\approx$0.74\,s \\[4pt]
\midrule
\multicolumn{4}{l}{\textit{SHAP-OSP design (one-time per domain) — PLIF, $128{\times}32$, 1 ch.}} \\[2pt]
$n_{\text{coali}}=430$  ($1\times$) & block $1\times1$ & 1$\times$ A40 & $\approx$15\,min \\
$n_{\text{coali}}=858$  ($2\times$) & block $1\times1$ & 1$\times$ A40 & $\approx$29\,min \\
$n_{\text{coali}}=2145$ ($5\times$) & block $1\times1$ & 1$\times$ A40 & $\approx$1\,h\,9\,min \\
$n_{\text{coali}}=3000$ ($7\times$) & block $1\times1$ & 1$\times$ A40 & $\approx$1\,h\,36\,min \\
$n_{\text{coali}}=3000$             & block $2\times2$ & 1$\times$ A40 & $\approx$1\,h\,36\,min \\
$n_{\text{coali}}=3000$             & block $4\times4$ & 1$\times$ A40 & $\approx$1\,h\,36\,min \\[4pt]
\multicolumn{4}{l}{\textit{SHAP-OSP design (one-time per domain) — DNS, $288{\times}96$, 2 ch.}} \\[2pt]
$n_{\text{coali}}=3000$ ($7\times$) & block $1\times1$ & 1$\times$ A100 & $\approx$5\,h \\
\bottomrule
\end{tabular}
\end{table}

Several aspects of this cost profile are noteworthy. MAPGA inference per channel is approximately $8\times$ slower on the DNS dataset than on the PLIF dataset (6\,s vs.\ 0.74\,s), consistent with the ${\sim}6.75\times$ larger grid ($288{\times}96$ vs.\ $128{\times}32$). The same grid-size scaling applies to SHAP-OSP: each coalition evaluation is one MAPGA call, so the DNS SHAP budget ($n_{\text{coali}}=3000$) costs ${\approx}5$\,h per channel per seed on an A100, compared to ${\approx}1$\,h\,36\,min per channel per seed on an A40 for PLIF; accounting for the hardware difference (A100 vs.\ A40), the per-channel SHAP cost is again broadly consistent with the grid-size ratio.

Within the PLIF ablation, wall-clock scales near-linearly with coalition count (ratios $430\,{:}\,858\,{:}\,2145\,{:}\,3000 \approx 1\,{:}\,1.9\,{:}\,4.6\,{:}\,6.4$), confirming MAPGA inference as the dominant cost per coalition. Segmentation granularity does not affect wall-clock time at fixed coalition budget: block sizes $1{\times}1$, $2{\times}2$, and $4{\times}4$ all require ${\approx}1$\,h\,36\,min, establishing coalition count, not feature dimensionality, as the sole bottleneck. As discussed in Supplementary Note~4, the practical benefit of coarser segmentation lies in requiring fewer coalitions for convergence, not in reducing per-coalition cost.

MAPGA reconstruction at deployment is fast (${\approx}0.74$\,s per channel on an A100 for the PLIF grid), enabling near-real-time inference once the sensor layout is fixed. SHAP-OSP design is a one-time offline investment per deployment domain, fully parallelisable as independent single-GPU array tasks. Users under tight compute budgets may adopt the $1\times$ coalition configuration ($n_{\text{coali}}=430$), achieving comparable OSP quality at a $6\times$ reduction in wall-clock time, as confirmed by the coalition-size ablation in Supplementary Note~4.

\section*{Supplementary Note 6. Discussion on temporal coherence and spatiotemporal extensions}

The diffusion prior used in Diff–SPORT is trained on single flow snapshots sampled independently from statistically stationary turbulent simulations. While this prior does not explicitly enforce trajectory-level temporal coherence, it learns the stationary distribution of the flow field, which implicitly encodes space–time statistics of the underlying dynamical system under the ergodicity assumption commonly adopted in turbulence modeling.

In statistically stationary urban turbulence, such snapshot-based learning is sufficient for a broad class of sensing and reconstruction tasks, where the objective is to recover instantaneous flow states or their ensemble statistics from sparse observations, rather than to forecast long-horizon trajectories. Extreme non-stationary regimes or distributional shifts beyond the training data are therefore considered out of scope in the present work.

\begin{figure}[htbp]
    \centering
    \begin{subfigure}[t]{0.55\textwidth}
        \centering
        \includegraphics[width=\textwidth]{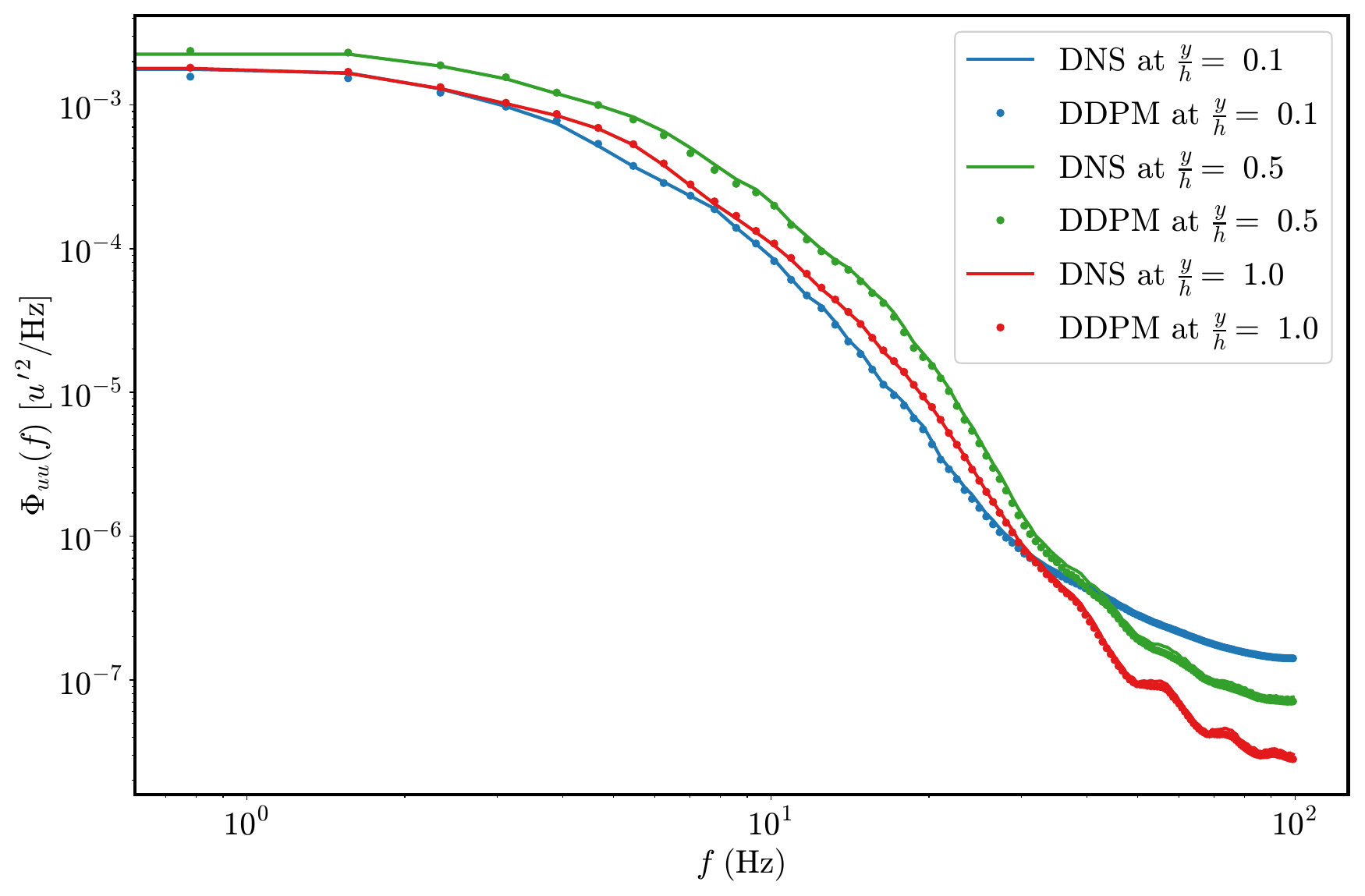}
        \caption{Power spectral density of streamwise velocity fluctuations at multiple wall-normal locations.}
        \label{fig:psd_u}
    \end{subfigure}
    \hfill
    \begin{subfigure}[t]{0.55\textwidth}
        \centering
        \includegraphics[width=\textwidth]{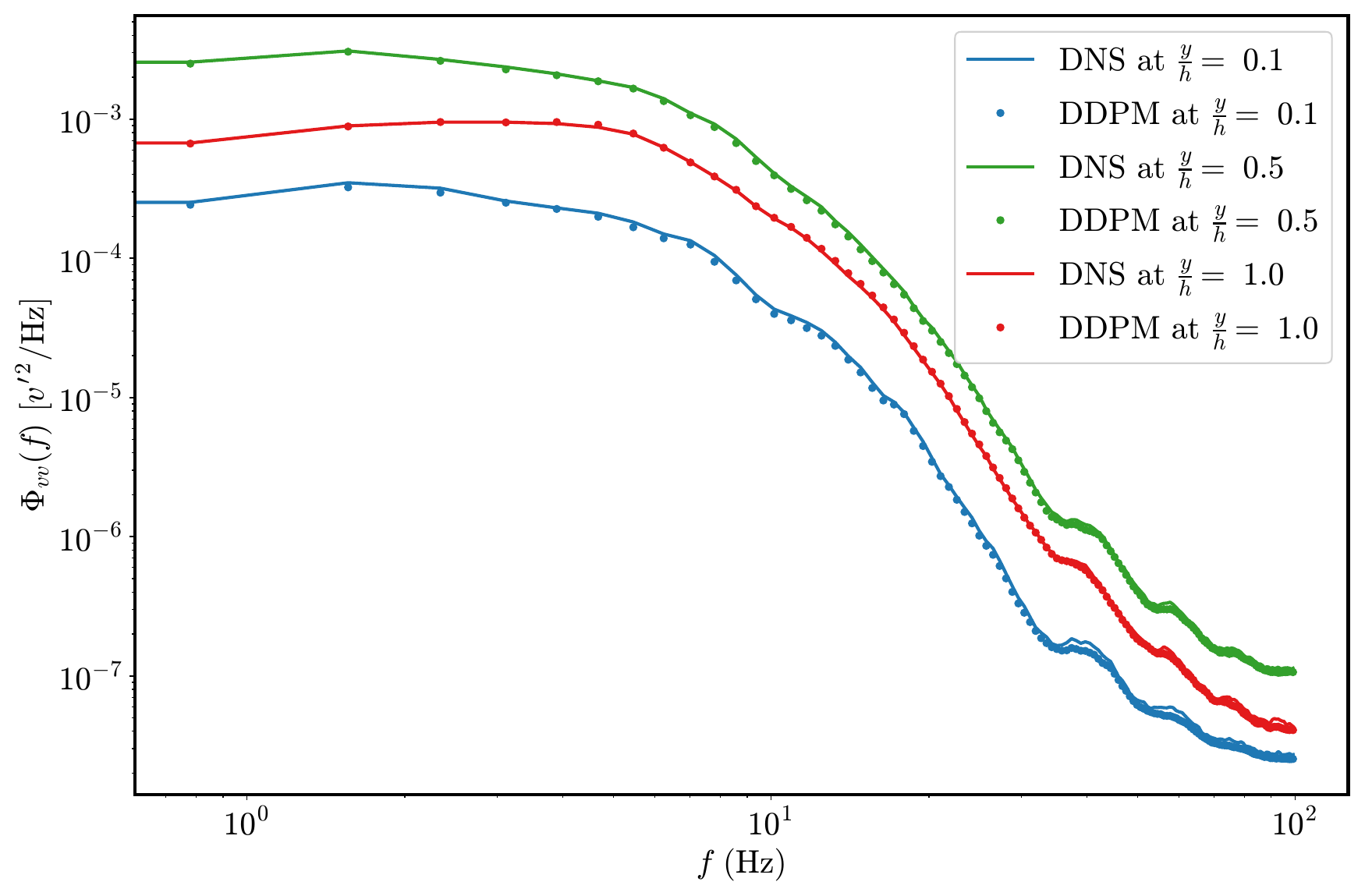}
        \caption{Power spectral density of wall-normal velocity fluctuations at multiple wall-normal locations.}
        \label{fig:psd_v}
    \end{subfigure}
    \caption{
    Comparison of power spectral densities (PSD) computed from unconditional samples of the Diff--SPORT diffusion prior and DNS reference data. The agreement across frequencies demonstrates that, despite IID snapshot training, the learned prior reproduces the temporal energy distribution of the statistically stationary turbulent flow. Source data are provided as a Source Data file.
    }
    \label{fig:psd_temporal_coherence}
\end{figure}

To assess whether the learned snapshot prior preserves relevant temporal statistics, we evaluate the power spectral density (PSD) of velocity fluctuations computed from unconditional samples of the trained diffusion model. As shown in Supplementary Fig. \ref{fig:psd_temporal_coherence}, the PSDs of generated samples closely match those of the DNS reference data across multiple wall-normal locations and frequency ranges. This agreement indicates that, despite the IID training formulation, the learned prior reproduces the correct distribution of temporal energy across scales, providing statistical temporal coherence rather than explicit trajectory continuity.

This design choice intentionally differs from recent spatiotemporal diffusion frameworks, such as CoNFiLD \cite{du_confild_2024}, which learn diffusion priors over entire time sequences and explicitly generate temporally coherent trajectories. While sequence-based priors are well suited for long-horizon synthesis and autoregressive generation, the IID formulation adopted here provides greater flexibility for inverse problems conditioned on sparse and irregular sensor measurements, where generating full space--time sequences may be unnecessary or computationally inefficient. In particular, the snapshot-based conditioning strategy avoids the need for temporally synchronized measurements and naturally accommodates irregular sampling in time, which is common in real-world urban sensing deployments. This flexibility is expected to facilitate scalability to large, city-scale sensor networks without imposing additional constraints on temporal data availability.

Importantly, the Diff–SPORT framework is modular with respect to the choice of prior. The proposed MAPGA posterior inference and SHAP-based sensor placement components are agnostic to whether the underlying diffusion prior is snapshot-based or spatiotemporal. Replacing the current IID prior with a spatiotemporal latent diffusion model would therefore preserve the overall pipeline while enabling explicit temporal coherence, at the cost of increased data, memory, and training complexity.

Extending Diff–SPORT to non-stationary or strongly transient flow regimes would require learning priors conditioned on evolving flow statistics, external covariates, or regime-dependent latent variables. Addressing such challenges offer important directions for future work but lies beyond the scope of the present study.

\section*{Supplementary Note 7. Reconstruction of fluctuating versus mean velocity fields}

Many fluid-mechanical analyses, including force and vorticity computations, are formulated in terms of the instantaneous (true) velocity field. In this work, Diff--SPORT focuses on reconstructing the fluctuating velocity component rather than jointly inferring both the mean and fluctuating fields. This is a deliberate and principled modeling choice aligned with turbulence theory and the statistical assumptions of the problem. In this section we will briefly describe why it is a natural choice to predict fluctuations over full-flow fields for sparse reconstruction problems.

For statistically stationary turbulent flows, the velocity field admits the Reynolds decomposition (Eq.~(1)),
$\mathbf{\Psi}(\mathbf{x},t)=\overline{\mathbf{\Psi}}(\mathbf{x})+\mathbf{\Psi}'(\mathbf{x},t)$.
In this study, the mean field $\overline{\mathbf{\Psi}}$ is removed from the three-dimensional DNS data during preprocessing, and the diffusion prior is trained exclusively on the fluctuating component $\mathbf{\Psi}'$. Accordingly, all reconstructions produced by Diff--SPORT correspond to $\mathbf{\Psi}'$, consistent with the stationarity assumption.

At inference time, Diff--SPORT operates on sparse instantaneous measurements of the form
\begin{equation}
\mathcal{S} = H \odot \mathbf{\Psi}(\cdot,t_0) + \boldsymbol{\eta}
           = H \odot \overline{\mathbf{\Psi}} + H \odot \mathbf{\Psi}'(\cdot,t_0) + \boldsymbol{\eta},
\label{eq:obs}
\end{equation}
where $H$ denotes a binary spatial mask defining the sensor locations, $\odot$ is the elementwise product, and $\boldsymbol{\eta}$ represents measurement noise. Given such sparse observations at a single time instant $t_0$, the decomposition in Eq.~(1) is fundamentally non-identifiable.

Indeed, for any spatial field $\boldsymbol{\delta}(\mathbf{x})$ that is unobserved by the sensor mask, i.e.,
\begin{equation}
H \odot \boldsymbol{\delta} = \mathbf{0},
\end{equation}

one may define an alternative decomposition,
\begin{equation}
\overline{\mathbf{\Psi}}^{\,*} = \overline{\mathbf{\Psi}} + \boldsymbol{\delta}, \qquad
\mathbf{\Psi}^{\prime *}(\cdot,t_0) = \mathbf{\Psi}'(\cdot,t_0) - \boldsymbol{\delta},
\end{equation}

which produces the same measurements:
\begin{equation}
H \odot \overline{\mathbf{\Psi}}^{\,*} + H \odot \mathbf{\Psi}^{\prime *}(\cdot,t_0)
= H \odot \overline{\mathbf{\Psi}} + H \odot \mathbf{\Psi}'(\cdot,t_0).
\end{equation}

Because sparse sensor masks $H$ leave large portions of the domain unobserved, infinitely many pairs $(\overline{\mathbf{\Psi}}, \mathbf{\Psi}')$ are consistent with the same instantaneous sparse observations. Moreover, the defining constraint $\mathbb{E}_t[\mathbf{\Psi}']=\mathbf{0}$ is inherently temporal and cannot be enforced from a single snapshot without long-time averaging or ensemble information. Consequently, jointly inferring the spatially varying mean field and instantaneous fluctuations from sparse instantaneous data is a fundamentally ill-posed inverse problem, independent of the specific learning architecture employed.

In highly turbulent flows, particularly those generated by DNS, the fluctuating component is high-dimensional, stochastic and dominates uncertainty in many quantities of interest, whereas the mean field is comparatively smooth and easier to estimate. In practice, mean velocity fields are commonly available from long-time sensor averages, baseline numerical simulations (e.g., RANS or LES), historical climatologies or can be simply predicted using simpler neural networks. Diff--SPORT therefore targets the more challenging and information-rich component $\mathbf{\Psi}'$. In the present work, the assumption of statistical stationarity implies that the training mean closely aligns with the mean of the test data. More generally, Diff--SPORT remains fully compatible with externally supplied mean fields when reconstruction of the full velocity field $\mathbf{\Psi}$ is desired. This separation between mean and fluctuations follows established practice in both fluid dynamics and recent diffusion-based modeling efforts. Notably, the CorrDiff framework adopts an explicit two-step strategy in which a deterministic model predicts the conditional mean and a diffusion model generates the residual, motivated by Reynolds decomposition and variance reduction arguments \cite{corrdiff}. Hence, it is imperative to note that learning the distribution of fluctuations is the central and most challenging task.

\section*{Supplementary Note 8. Field-dependence of SHAP sensor placement: velocity versus concentration on the same urban canopy}
\label{sec:shap_field_dependence}

The SHAP-based sensor placement reported for the LT2400 urban canopy (main text, concentration field $C'$) allocates the majority of high-importance sensors upstream of the first building, with comparatively little importance assigned within the wake of the downstream building cluster. This raises a natural question: does SHAP systematically undervalue the wake, or does the placement instead reflect a property of the governing physics specific to passive-scalar transport? We address this directly by repeating the full Diff--SPORT framework (diffusion prior, MAPGA, SHAP-OSP) on a velocity dataset, specifically streamwise velocity component ($u'$), acquired using particle image velocimetry (PIV) and comparing the resulting sensor placement to the concentration case in a similar environment. We also provide a theoretical explanation for this behavior below.

\subsection{Theoretical basis: elliptic versus parabolic information propagation}

The two fields obey governing equations of fundamentally different mathematical character, a fact that determines how information about the flow state propagates spatially and, consequently, which sensor locations carry the most non-redundant information for reconstruction.

The incompressible velocity field is coupled through the pressure Poisson equation,
\begin{equation}
\nabla^2 P = -\rho\, \nabla\!\cdot\!\left(\mathbf{U}\cdot\nabla\mathbf{U}\right),
\end{equation}
where $P$ denotes the pressure and $\rho$ the fluid density. This equation is elliptic: a perturbation anywhere in the domain, including the wake, influences the pressure (and hence velocity) field everywhere else, including upstream. Equivalently, in a two-dimensional slice the velocity follows from the vorticity via a stream-function Poisson equation, $\nabla^2\psi=-\omega$, where $\psi$ is the stream function and $\omega$ the vorticity, so a vortex shed in the wake induces velocity upstream through the Biot–Savart kernel. Wake vortex shedding is further governed by self-sustained, quasi-periodic dynamics rather than being fully determined by the upstream state, so wake measurements carry information about the velocity field that upstream sensors cannot supply.

The passive scalar, in contrast, obeys the advection--diffusion equation,
\begin{equation}
\frac{\partial C}{\partial t} + \underbrace{\mathbf{U}\cdot\nabla C}_{\text{advection}} = \underbrace{D\nabla^2 C}_{\text{diffusion}},
\end{equation}
which is parabolic and, at high Péclet number, approaches hyperbolic behaviour: information propagates predominantly downstream along mean-flow characteristics. For LT2400 ($U_\infty=0.46$~m/s, $\delta=83$~mm, $Sc=2500\pm300$, $\nu\approx10^{-6}$~m$^2$/s), the molecular Péclet number is

\begin{equation}
Pe_{\text{mol}} = \frac{U_\infty\,\delta\,Sc}{\nu} \approx 9.5\times10^{7},
\end{equation}

i.e.\ advection overwhelmingly dominates molecular diffusion. Under this regime the downstream concentration is a forward function of the upstream inlet condition, determined entirely by what entered upstream at earlier times, while the inverse (inferring the upstream state from a downstream, already-mixed measurement) is ill-posed: turbulent mixing is thermodynamically irreversible and destroys information, so multiple upstream states can map to the same downstream mixed state.

For a SHAP-OSP procedure that quantifies marginal reconstruction utility given a learned prior $p_\theta$, this asymmetry directly predicts opposite placement strategies for the two fields: an upstream sensor constrains the (a priori unknown, snapshot-varying) scalar inlet condition that determines the entire downstream field, so upstream sensors attain large SHAP importance for $C'$; a wake sensor for $u'$ captures autonomous vortex-shedding information that is not recoverable from upstream measurements alone, so wake sensors retain non-negligible importance for $u'$ in a way they do not for $C'$. 

\subsection{Empirical test for varying field type}

The PIV experimental process mirrors the PLIF as closely as possible: acquired under similar facility and inflow conditions and restricted to the streamwise fluctuation component $u'$ for a single-channel, like-for-like comparison with $C'$, with the target resolution, normalization strategy, feasibility-region construction, and SHAP procedure ($n_{\text{coali}}=3000$ coalitions, pixel-level segmentation) all kept the same. We note that the PIV and PLIF acquisitions, while sharing the same campaign label and nominal geometry, were not acquired with pixel-identical building footprints in the processed data used here; the comparison below is therefore a same-campaign, same general urban-canopy class comparison rather than a literal pixel-overlay, and we restrict our claims accordingly to the qualitative placement pattern rather than a location-by-location match.

\begin{figure}[h]
    \centering
    \includegraphics[width=0.8\textwidth]{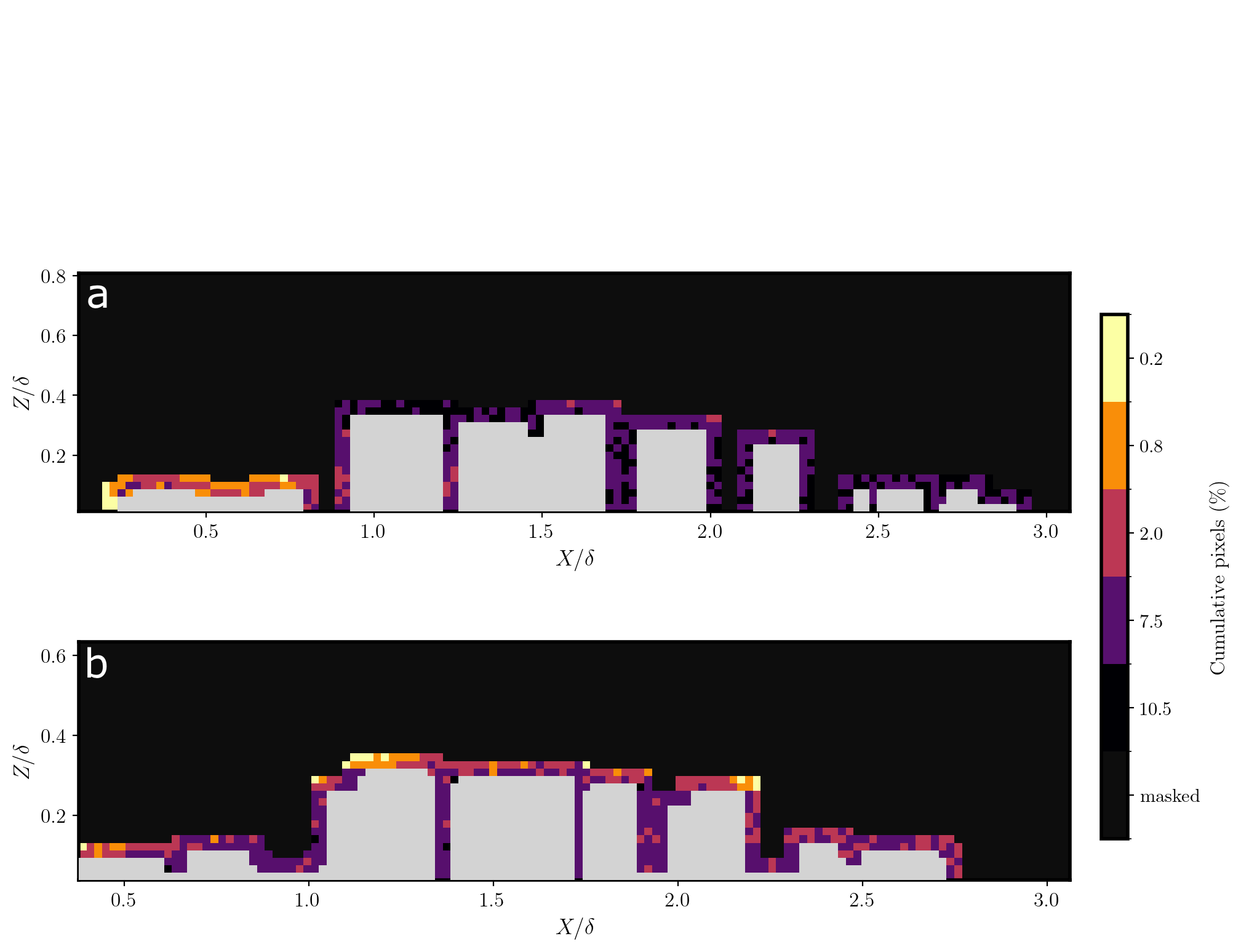}
    \caption{
    \textbf{SHAP sensor importance depends on field type on a shared urban-canopy campaign.}
    Cumulative SHAP importance maps, identical color scale and sensor-region construction ($n=2$-pixel building-boundary expansion), for (a) concentration $C'$ (PLIF) and (b) streamwise velocity $u'$ (PIV), acquired under similar experimental campaign. Both axes are normalized by the boundary-layer thickness $\delta=83$~mm in a shared coordinate frame. Grey regions denote buildings; black denotes valid flow outside the candidate sensor region (masked, not evaluated by SHAP). Panel (a) shows importance concentrated immediately upstream of the first building; panel (b) shows importance distributed around building surfaces across the full streamwise extent of the canopy, including downstream rooftops and edges. Building footprints differ slightly between the two acquisitions; the comparison is same-campaign, not a pixel-identical overlay.
    }
    \label{fig:shap_piv_vs_plif}
\end{figure}

Figure~\ref{fig:shap_piv_vs_plif} shows the resulting SHAP importance maps side by side, using an identical cumulative-importance color scale and identical sensor-region construction for both fields. For concentration (panel a), the highest-importance sensors concentrate immediately upstream of the first building, consistent with the main-text result. For velocity (panel b), the highest-importance sensors are instead distributed around building surfaces spanning the full streamwise extent of the canopy, including rooftops and edges of the downstream buildings, rather than concentrating upstream. This qualitative contrast, obtained with an unchanged framework and an unchanged sensor-region construction on the same facility, is precisely the behaviour predicted by the elliptic-versus-parabolic distinction above. We restrict this comparison to the spatial placement pattern; reconstruction-error magnitudes are not compared between fields, as $u'$ and $C'$ differ in physical units and normalization and are not meaningfully comparable on a common numerical scale.

\section*{Supplementary Note 9. Robustness of zero-shot inference under geometric distribution shift: an unretrained prior on an unseen canopy geometry}

The main text demonstrates task-level zero-shot inference: a single fixed diffusion prior, trained once per physical domain, supports arbitrary downstream reconstruction and sensor-placement queries without retraining. Here we assess, in a preliminary manner, the capability of the framework when the prior is applied to a physical domain it was never trained on. This test is primarily included to characterize the behaviour of the inference pipeline under a controlled distribution shift.

\begin{figure}
    \begin{center}
        \includegraphics[width=0.8\textwidth]{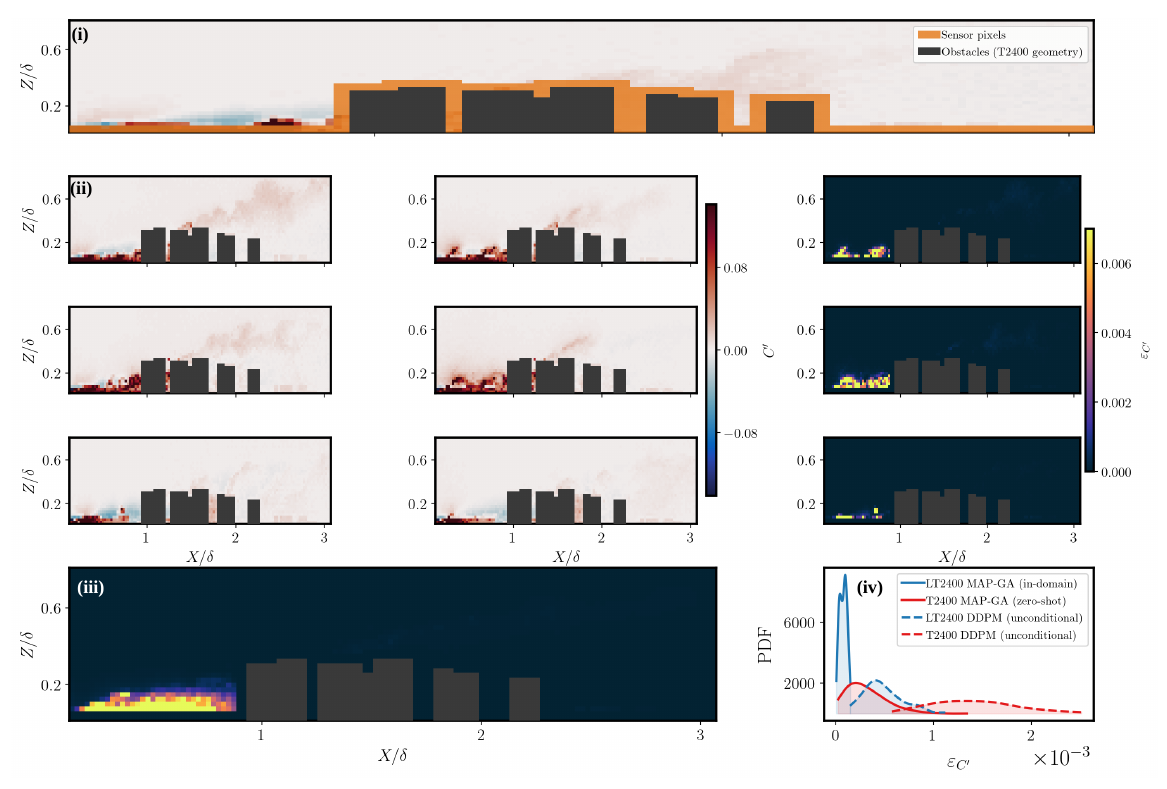}
        \caption{\label{fig:zeroshotinferencet2400}
        \textbf{Zero-shot MAPGA reconstruction on the unseen T2400 canopy
        geometry using the unretrained LT2400 prior.}
        (i) feasibility mask applied to a representative snapshot, with sensor
        pixels constructed on the T2400 obstacle geometry;
        (ii) three randomly selected test snapshots each displaying ground-truth
        concentration ($C^{\prime}$, left), zero-shot MAPGA reconstruction
        (centre) and instantaneous absolute error ($\varepsilon_{C^{\prime}}$,
        right);
        (iii) mean absolute error field ($\varepsilon_{C^{\prime}}$) averaged
        over the full T2400 test dataset;
        (iv) error PDFs comparing MAPGA (solid lines) and unconditional DDPM
        (dashed lines) for LT2400 and T2400. The identical
        checkpoint, sensor mechanism and inference algorithm are used
        throughout, with no retraining or recalibration on T2400. Source data are provided as a Source Data file.
        }
    \end{center}
\end{figure}

We applied the unmodified, unretrained LT2400-trained DDPM prior to the T2400 dataset. This dataset consists of PLIF concentration measurements acquired in the same facility under the same inflow conditions as LT2400, but over a physically distinct canopy geometry containing only the tall buildings (Figure~\ref{fig:zeroshotinferencet2400}, top panel) never seen during training. The T2400 concentration field is statistically distinct from the training distribution: both its mean and variance are substantially higher than LT2400's. We deliberately retained LT2400's normalization and mean field without recalibration, so the comparison reflects a genuinely unadapted prior applied to an out-of-distribution field. The feasibility mask was constructed on the T2400 geometry following the same procedure as in the main text (ground level and building facades) while the sensor mechanism, MAPGA inference algorithm, and checkpoint are identical to those used for LT2400 throughout.


Figure~\ref{fig:zeroshotinferencet2400} compares zero-shot MAPGA reconstruction against unconditional DDPM generation for both datasets using this same prior, isolating the contribution of sensor conditioning and domain fit. MAPGA reconstruction on T2400 shifts the error distribution well below the T2400 unconditional baseline, and also below the unconditional baseline of the training dataset (LT2400) itself, mirroring the general pattern observed in-domain. As expected we observe that zero-shot reconstruction degrades moderately relative to in-domain MAPGA and the distribution is broader. The dominant errors concentrate in the near-inlet region upstream of the first building, where the T2400 concentration statistics depart most strongly from the training distribution, but these results show great promise for future geometry-generalization work.

\bibliography{Diff-SPORT-diffusion}